\documentclass[letterpaper]{article} %
\usepackage{aaai25}  %
\usepackage{times}  %
\usepackage{helvet}  %
\usepackage{courier}  %

\usepackage{booktabs}
\usepackage{subcaption}
\usepackage{cleveref}

\usepackage[hyphens]{url}  %
\usepackage{graphicx} %
\usepackage{natbib}  %
\usepackage{caption} %
\usepackage{algorithm}
\usepackage{algorithmic}
\usepackage{fontawesome5}

\usepackage{newfloat}
\usepackage{listings}
\DeclareCaptionStyle{ruled}{labelfont=normalfont,labelsep=colon,strut=off} %
\floatstyle{ruled}
\newfloat{listing}{tb}{lst}{}
\floatname{listing}{Listing}
\title{Characterizing Bluesky Content Moderation Service:\\From Automation of Service to Landscape of Harms}
\author{
    Pushpdeep Singh\textsuperscript{\rm 1}\thanks{Corresponding author, psingh@mpi-sws.org},
    Sayeh Jarollahi\textsuperscript{\rm 2},
    Ayan Majumdar\textsuperscript{\rm 1,2},
    Vabuk Pahari\textsuperscript{\rm 1},
    Abhijnan Chakraborty\textsuperscript{\rm 3},
    Krishna P. Gummadi\textsuperscript{\rm 1},
    Ingmar Weber\textsuperscript{\rm 2},
    Abhisek Dash\textsuperscript{\rm 1}
}
\affiliations{
    \textsuperscript{\rm 1}Max Planck Institute for Software Systems (MPI-SWS), Germany\\
    \textsuperscript{\rm 2}Saarland University, Germany\\
    \textsuperscript{\rm 3}Indian Institute of Technology Kharagpur, India
}

\usepackage{bibentry}

\newcommand{\new}[1]{\textcolor{black}{#1}}

\newcommand{\lintolerant}{\textit{intolerant}}
\newcommand{\lrude}{\textit{rude}}
\newcommand{\lthreat}{\textit{threat}}
\newcommand{\lspam}{\textit{spam}}
\newcommand{\lselfharm}{\textit{self-harm}}
\newcommand{\lgraphicmedia}{\textit{graphic-media}}
\newcommand{\lporn}{\textit{porn}}
\newcommand{\lsexual}{\textit{sexual}}
\newcommand{\lnudity}{\textit{nudity}}
\newcommand{\lhide}{\textit{!hide}}
\newcommand{\lwarn}{\textit{!warn}}
\newcommand{\ltakedown}{\textit{!takedown}}
\newcommand{\lsexualfigurative}{\textit{sexual-figurative}}

\usepackage{xcolor}

\newcommand{\neww}[1]{\textcolor{black}{#1}}

\begin{document}
\nocopyright
\maketitle

\begin{abstract}

Empirical research on content moderation is fundamentally constrained by the opaque deployment of moderation systems on major social media platforms. 
To this end, the recent emergence of decentralized platforms with transparent, public moderation logs presents an unprecedented opportunity for independent audits. 
\new{In this work, we leverage this architectural transparency to conduct the first large-scale audit of the default moderation system on Bluesky, the \textit{Bluesky Moderation Service} (BMS). 
Analyzing its 10.6M moderation labels from 2025, we investigate three foundational aspects: (i)~its mechanism (the degree of automation versus human oversight), (ii)~its efficacy (accuracy in detecting harms), and (iii)~its purpose (the landscape of harms it identifies).}

Our findings reveal a human-AI collaborative system where labels for sexual and graphic content are applied automatically in seconds, while nuanced and high stakes labels require more human oversight, taking hours or days. 
\new{Through a manual annotation study, we find the BMS operates with high precision (0.837), but struggles with low recall (0.222), with our annotators identifying 4.5$\times$ more harmful content than the moderation system in a random sample. }
Finally, unsupervised clustering of the most frequently applied labeled posts uncovers detected harms ranging from hostility in discourse toward protected groups to the spread of sexually explicit and other graphic content.  
Our work offers a look into the operational realities of a deployed moderation system, providing a concrete data-driven foundation for designing more effective and transparent moderation systems.
\footnote{\textcolor{red}{This work has been accepted for publication at ICWSM 2027.}}

\end{abstract}

\begin{links}
  \link{Code \faGithub}{https://github.com/iampushpdeep/BMS}
  \link{Project \scalebox{0.85}{\faGlobeAmericas}}{https://bms-audit.github.io}
\end{links}

\maketitle

\section{Introduction}
\label{Sec: Intro}
Social media platforms deploy content moderation systems to govern the discourse on their forums. 
The stakes of such moderation are immense: while under-moderation may allow unchecked spread of unsafe content resulting in systemic harms, over-moderation risks stifling legitimate expression, leading to censorship~\cite{gillespie2018custodians}. 
Given this delicate balance, it is imperative to audit the functioning of deployed moderation systems on social media platforms. 

However, independent audit of content moderation systems has been fundamentally constrained by the \textit{non-transparent deployment} on established social media platforms~\cite{gillespie2018custodians, suzor2019we, juneja2020through, roberts2016commercial}.  
Specifically, their moderation policies, decision-making algorithms, and enforcement data are proprietary: shielding them from independent scrutiny. 
In fact, such a lack of transparency has led policymakers to adopt regulations requiring greater transparency obligations for deployed content moderation systems. 
Concretely, the Digital Services Act (DSA) in the EU mandates platforms to disclose their moderation practices, including the degree of human oversight, the use of automated tools and the rationale behind moderation actions~\cite{EU2022DSA}. 
However, even after these efforts, our understanding of prevalent systemic harms %
is limited due to \textit{non-transparent reporting} by platforms~\cite{trujillo2025dsa, shahi2025year}.

A new opportunity in this landscape arises from the emergence of decentralized platforms built on open protocols~\cite{kleppmann2024bluesky, raman2019challenges, Balduf2024LookingATA}. 
By design, decentralized platforms offer greater transparency into their operations. 
For example, Bluesky~({\tt https://bsky.app}), built on the Authenticated Transfer Protocol, publishes its moderation actions as a publicly accessible data stream~\cite{kleppmann2024bluesky}. 
Such architectural transparency transforms content moderation from an inscrutable `black box' into an \textit{auditable transparent public log}. 
\new{This improved transparency enables some foundational inquiry into the \textit{mechanism}, \textit{efficacy} and \textit{purpose} of the deployed content moderation systems.}

Leveraging the publicly accessible moderation logs of Bluesky's default moderation system (known as the \textit{Bluesky Moderation Service} (BMS)\footnote{\texttt{https://bsky.app/profile/moderation.bsky.app}}), in this study, we aim to answer the following important questions:

\vspace{0.5 mm}
\noindent
\textbf{RQ1: }\textit{What is the degree of automation and human oversight in the Bluesky Moderation Service?
}\\
\noindent
\textbf{RQ2: }\new{\textit{How effective is the Bluesky Moderation Service 
in detecting harms on the platform?
}}\\
\noindent
\textbf{RQ3: }\textit{What is the landscape of harms detected by the Bluesky Moderation Service 
on the platform?}
\vspace{0.5 mm}

\noindent
\textbf{Why are these questions important?} 
Increasingly content moderation relies on \textit{human-AI} collaborations~\cite{Bluesky2025TOS, gillespie2018custodians, halevy2022preserving}. 
To this end, the primary question regarding the moderation mechanism centers on the interplay between scalable automated detection and nuanced human oversight or interventions.  
\new{While a number of platforms report such information through their transparency reports, due to the lack of procedural and outcome-level transparency, such details are rarely verifiable. 
RQ1 moves beyond platform assurances to empirically quantify the degree of automation and human oversight, providing a verifiable blueprint of the system's underlying mechanism.} %

\new{An understanding of the mechanism is insufficient without a rigorous evaluation of its performance, i.e., \textit{how effective is the mechanism in practice}? 
While platforms may self-report high performance, such metrics are calculated on proprietary data and are impossible to independently validate. 
By assessing precision (what fraction of the detected content is actually harmful) and recall (what fraction of all harmful content has been detected), we aim to quantify the system's real-world effectiveness.
}

Beyond the moderation mechanism and its effectiveness, another line of inquiry aims at the \textit{purpose} of moderation systems by characterizing the landscape of harms it identifies. 
Such characterization is crucial for two reasons: (a)~it reveals how closely stated principles, i.e., \textit{moderation policies}, align with their {\it de facto} operationalization: {\it moderation practices}; and (b)~it offers a ground-level view of the systemic harms present on the platform.

To answer these questions, we conduct the first large-scale audit of the default content moderation system deployed on Bluesky. %
It is applied by default to every user upon signing up on Bluesky, \new{thereby establishing a universal safety standard for nearly 40M users on the entire platform~\cite{Bluesky2025Report}. 
We gathered and analyzed labels assigned by the BMS for $10.6$M posts during 2025.} 
Our analyses reveal the following key findings about BMS:  

\noindent
$\bullet$\textbf{ Findings for RQ1:} By using labeling delay as a proxy, our analyses show that the BMS labels sexual and graphic content quickly (e.g., median 7.1 seconds for \lporn{} label). 
Whereas for more nuanced labels, it relies more on human oversight, incurring longer delays (e.g., median 8.4 days for \lrude{} label). 
\new{Our findings validate the disclosures made in Bluesky's recent transparency report~\cite{Bluesky2025Report}}.

\noindent
$\bullet$\textbf{ Findings for RQ2:} \new{By annotating a set of 1000 labeled posts and 1000 random posts from Bluesky's firehose, we find that the BMS has high precision, but it suffers from low recall. 
While human annotators (coauthors) agree with 83.7\% of all posts labeled by BMS in the set of 1000 labeled posts (precision 0.837), they flag 4.5$\times$ more posts to be harmful in the 1000 random posts (recall 0.222). }

\vspace{0.5 mm}
\noindent
$\bullet$\textbf{ Findings for RQ3:} By deploying unsupervised clustering on the labeled posts, we find several granular categories (within each label) of harmful content detected and moderated by BMS. 
While BMS detects more sexually explicit content, the most taken-down posts contain violent death wishes to political leaders. 
Our analyses reveal that many labels share overlap among the categories, which could lead to challenges for moderators and confusion for end-users.

\section{Related Work}
\label{Sec: Related}
\neww{We position our work within three strands of prior work and provide a brief overview of each: (a)~content moderation, (b)~regulatory intervention, and (c)~decentralized platforms.}

\subsection{Content Moderation}
Content moderation has become a crucial component of social media platforms, which often mediate how consumers communicate and consume information online. 
The Digital Services Act (DSA) defines content moderation as \textit{activities undertaken by online platforms aimed at detecting, identifying and addressing information incompatible with their terms and conditions}~\cite{EU2022DSA}. 
Thus, it aims to strike a delicate balance between ensuring safety and empowering free speech~\cite{gillespie2018custodians}.
Given its importance, a number of prior works have investigated content moderation on deployed online platforms. 
While some works study moderation mechanisms and their drawbacks~\cite{halevy2022preserving, Hartmann2025LostIMA, roberts2019behind, qiwei2024reportingnonconsensualintimatemedia, Hartmann2024WatchingTWA, ma2022m}, some others delve into the transparency desiderata of moderation systems~\cite{gillespie2018custodians, suzor2019we, juneja2020through, roberts2016commercial}. 
While valuable, the scope of such studies has often been constrained by the inherent opaqueness of traditional centralized platforms, making it difficult to comprehensively map the landscape of prevalent systemic harms and the corresponding moderation actions.

\subsection{Regulatory Intervention}
To improve transparency of centralized moderation systems, DSA requires all online platforms to submit statements of reasons for all undertaken moderation actions. 
This has resulted in the creation of the DSA Transparency Database~\cite{europaHomeTransparency}. %
However, several recent studies showcase the lack of compliance and inadequacies in the transparency database, ranging from limited reporting from platforms to a lack of information about content being moderated~\cite{trujillo2025dsa, shahi2025year}. %
These inadequacies further limit the ability of researchers to understand the current state of systemic harms in digital platforms and the effectiveness of the detection strategies. %

\subsection{Decentralized Platforms}
In this context, decentralized social media platforms (e.g., Mastodon, Bluesky) deployed on open protocols (e.g., Activity Pub, Authenticated Transfer Protocol) have emerged as more transparent alternatives, triggering new opportunities for public interest research. 
These platforms, deployed on open protocols, seem to be inherently more auditable than centralized platforms, which are increasingly becoming walled gardens by further restricting data access. 
Such data access has resulted in several large-scale studies on Mastodon~\cite{raman2019challenges, zhang2024trouble, bono2024exploration}, and Bluesky~\cite{kleppmann2024bluesky, Balduf2024LookingATA, balduf2025bootstrapping} in recent years. 
\new{~\citet{Failla_2024} released a large-scale dataset of Bluesky user activity, covering over 4M accounts and 235M posts. ~\citet{nogara2025longitudinalanalysismisinformationpolarization} conduct a longitudinal analysis of user activity around Bluesky's public launch, examining misinformation dynamics, political polarization, and toxicity levels.}

\new{~\citet{Balduf2024LookingATA} provides a foundational contribution to the study of the Bluesky platform, and conducts the first large-scale architectural analysis of the platform and its sub-components: Labelers, Feed Generators, and Personal Data Servers, and provides an initial glimpse into the labeling ecosystem and third-party moderation providers. However, their focus remains infrastructural; the operation and evaluation of moderation decisions and the specific harms the labeler service addresses fall outside their scope of work.}

To the best of our knowledge, this is the first study to perform a large-scale, empirical characterization of a transparent moderation system, analyzing its operational mechanisms, efficiency, and the specific harms it addresses.

\section{Bluesky and its Content Moderation}
\label{Sec: Background}

Before we answer the key research questions, in this section we first provide a brief overview of the Bluesky ecosystem, its key architectural components, and its default moderation system, the Bluesky Moderation Service (BMS).

\vspace{1mm} \noindent
\textbf{Identity and Userbase on Bluesky: }
Bluesky is a decentralized social media platform built on the Authenticated Transfer protocol~\cite{kleppmann2024bluesky}. 
When a user signs up on Bluesky, they are assigned a unique \textit{Decentralized Identifier} (DID). 
Each DID can be resolved to a corresponding DID document maintained in the \textit{Public Ledger of Credentials} (PLC) directory. 
To quantify Bluesky's user base, we gathered the entire PLC directory using its `Export' endpoint (\texttt{https://web.plc.directory/api/redoc\allowbreak\#operation/Export}).
We observe a total of $66.9$M unique DID documents in the PLC directory. 
Such a large base of registered DIDs indicates the growing popularity of Bluesky and the need to study such evolving social media platforms. 
DID documents store information about where the corresponding user's data repository resides, their user handle (human-readable identifier), etc.
These DID documents are foundational to Bluesky's architecture, acting as a \textit{discovery mechanism} for some of its main components.

\subsection{Key Architectural Components in Bluesky}

In Bluesky, the different core functionalities %
are handled by separate architectural components~\cite{kleppmann2024bluesky}, namely: 
(a)~Personal Data Servers, (b)~Relay, (c)~Labeler, (d)~Feed generator, and (e)~App View.

\vspace{1mm} \noindent
\textbf{Personal Data Server (PDS):} 
A personal data server authenticates users registered to it and hosts their data.
Every Bluesky user must be registered with a single PDS. 
The service endpoint in a user's DID document, specifically the \texttt{\#atproto\_pds} entry, points to the PDS that hosts their repository. 
This repository includes records of all posts, likes, comments, and other actions taken by a user.

\vspace{1mm} \noindent
\textbf{Relay: } Relay aggregates newly created user records across all active PDSes in the network to create a stream of records called \textit{Firehose} that is publicly accessible. 

\vspace{1mm} \noindent
\textbf{Labeler: } Labeler is the central content moderation engine on Bluesky. 
Labelers function like a typical user account, and become a labeler by adding an \texttt{\#atproto\_labeler} service entry to their DID document. %
The labeler then publishes a \texttt{app.bsky.labeler.service} record in their repository to define and share their moderation policies.
The labeler consumes the firehose as input and labels selected incoming records to generate a \textit{label stream}---a publicly accessible log of actions taken by the labeler. 
It can apply labels to content, accounts, or profiles. 
The applied labels are leveraged to filter content and perform other moderation operations during content recommendation and dissemination.

\vspace{1mm} \noindent
\textbf{Feed generator: }Feed generator is the content recommendation arm of Bluesky. 
Any user can run their own feed generator that consumes the firehose and produces a relevant feed consisting of URIs pointing to filtered posts.

\vspace{1mm} \noindent
\textbf{App View: }The App View collates and indexes the data produced across the architecture by consuming the firehose and provides the results to users. 
In the context of content moderation, App View enforces the moderation labels on the user interface by taking the recommended actions. 

\subsection{Content Moderation on Bluesky}%
\label{sec: BMS}
Primarily, labelers consume the firehose and label content that satisfies certain criteria, e.g., content violating terms of service or belonging to specific categories such as nsfw or disturbing imagery. 
Alternatively, Bluesky users can report posts and accounts to labelers, who can label them based on those reports. 
Thus, labelers are the core component responsible for content moderation on Bluesky.

\noindent
\textbf{Bluesky Moderation Service: }To this end, labelers deployed by Bluesky PBC (BMS and country-specific moderation services) are applied to user accounts by default. 
While the open architecture of AT Protocol allows any third party to set up their labeler services, users are actively required to subscribe to them, which can naturally limit their reach. 
Hence,  we leave the study of how third-party labelers complement or contradict the BMS for future work. 
While most of the analyses we propose here can be extended to any transparent moderation systems deployed on AT protocol, for this study, we primarily focus on the BMS.

\noindent
\textbf{BMS' Content Moderation Policies and Labels: }
The policies %
associated with a labeler can be retrieved using the \texttt{app.bsky.\allowbreak labeler.service/self} record from the PDS using the \texttt{com.atproto.repo.get\allowbreak  Record} endpoint. 
For the BMS labeler, we collect these policies, which contain the list of labels they apply, their definitions, along with other attributes associated with each label, such as its \textit{default action} initially determined by the labeler. 
\Cref{tab:all_label_values} provides a summary of these label values, including their policy descriptions and default actions. 

BMS' labels include \lintolerant{}, \lrude{}, \lporn{}, \lsexual{}, \lspam{}, \ltakedown{}, etc. 
These labels can be broadly divided into two categories: (a)~\textit{harm labels}: refers to a certain category of harm (e.g., \lintolerant{}, \lporn{}, \lselfharm{}, etc.), (b)~\textit{action labels}: refers to an action taken by the Bluesky Moderation Service (e.g., \ltakedown{}, \lwarn{}, or \lhide{}).

Crucially, the \textit{default action} for each label, typically \textit{hide}, \textit{warn}, or \textit{ignore}, determines how labeled content is presented to a user who has not explicitly altered their preferences. 
\textit{Hide} configuration hides posts with that label from the user's feeds, \textit{warn} provides a warning before the content is viewed, and \textit{ignore} shows the label badge without any content filtering or visual obstruction. 
Based on the default action recommended by the labeler %
(or the modified setting of the user), the App View enforces the moderation labels to affect the Bluesky feed of the corresponding user.

\begin{table*}[t]
\centering
\caption{Labels used in our analysis, along with their names, descriptions, and their recommended default action. %
}
\label{tab:all_label_values}
\small
\setlength{\tabcolsep}{6pt} %
\resizebox{\textwidth}{!}{%
\begin{tabular}{l p{13cm} l}
\toprule
\textbf{Label} & \textbf{Policy Description} & \textbf{Default Action}\\
\midrule
\textbf{intolerant} & Discrimination against protected groups. & warn\\
\textbf{rude} & Rude or impolite, including crude language and disrespectful comments, without constructive purpose. & hide\\
\textbf{threat} & Promotes violence or harm towards others, including threats, incitement, or advocacy of harm. & hide \\
\textbf{sexual-figurative} & Art with explicit or suggestive sexual themes, including provocative imagery or partial nudity. & show\\
\textbf{porn} & Explicit sexual images. & hide \\
\textbf{sexual} & Does not include nudity. & warn \\
\textbf{nudity} & E.g., artistic nudes. & show \\
\textbf{graphic-media} & Explicit or potentially disturbing media. & warn \\
\textbf{self-harm} & Promotes self-harm, including graphic images, glorifying discussions, or triggering stories. & warn \\
\textbf{spam} & Unwanted, repeated, or unrelated actions that bother users. & hide \\
\textbf{!hide} & This content has been hidden by the moderators. & --\\
\textbf{!warn} & This content has received a general warning from moderators. & -- \\
\textbf{!takedown} & This content has been removed according to policy. & -- \\
\bottomrule
\end{tabular}}
\vspace{-4 mm}
\end{table*}

\vspace{1mm} \noindent
\textbf{Country-specific BMS: }Furthermore, there are country-specific labelers for Germany, India, Turkey, Russia, and Brazil, which by default apply when a user connects to Bluesky from the mentioned countries. 
These labelers are also run by Bluesky PBC. 
However, unlike other moderation services, the labels applied by these labelers are not specified in their service record. 
Furthermore, they also do not provide users with any choice to change the moderation settings. 
Bluesky PBC deploys these country-specific moderation services to comply with legal or regulatory requirements imposed by national authorities. 

Out of these five country-specific moderation services, only Germany, Turkey, and Brazil have issued labels. 
Turkey's labeler issued a total of 31 labels, consisting of 6 \textit{!hide} labels on \textit{posts} and 25 \textit{!hide} labels on \textit{accounts}. 
The Germany-specific labeler issued 68 labels, almost all of which were applied to posts (66 \textit{!hide} labels on \textit{posts}) and 2 \textit{!hide} labels on accounts.
The Brazil-specific labeler issued 16 labels, including 13 \textit{!hide} labels on posts, 1 \textit{!takedown} label on a post, and 2 \textit{!takedown} labels on accounts.

\subsection{Data Collection for Moderated Content}
\noindent
\textbf{Identifying BMS labeler: }
\new{Labelers can be identified via their DID documents. 
DIDs with an \texttt{\#atproto\_labeler} service entry are labelers. 
DID document also contains the \textit{serviceEndpoint} for the labeler, which can be used along with \texttt{com.atproto.label.subscribeLabels} API endpoint to retrieve the full label history available via the stream. We identified the \textit{serviceEndpoint} for the BMS %
(with DID \texttt{did:plc: ar7c4by46qjdydhdevvrndac}) as \texttt{mod.bsky.app}.
}%

\vspace{1mm} \noindent
\textbf{Bluesky's label stream data collection:} As %
the label stream (\texttt{com.atproto.\allowbreak label.subscribeLabels}) on AT protocol (output of the labelers) is publicly accessible, 
\new{to conduct an effective assessment of the BMS, we collected the label stream for the entire year of 2025. %
BMS applied $18,343,953$ labels in total, out of which $15,744,081$ %
were applied on posts, $2,418,534$ on accounts, $84,647$ on profiles, and $96,691$ on other record types (e.g., lists, feed generators).}%

\vspace{1mm} \noindent
\textbf{Additional metadata collection:} We further gathered the details of the records %
labeled by Bluesky. 
\new{The label stream only contains the Content Identifiers (CIDs) and AT URIs of labeled posts, not the complete post records (with metadata about when the post was created, embed information, post text, media content identifiers etc.). 
As an AT URI is of format \texttt{at://<did>/<collection>/<rkey>}, we use the DID to obtain the corresponding PDS endpoint via the PLC directory, and then retrieve the associated record using \texttt{com.atproto.repo.getRecord}. 
For records with media objects, we fetch the corresponding blob (image/thumb/video) using \texttt{com.atproto.sync. getBlob} endpoint}.

\new{%
We were able to gather post records corresponding to $10,681,824$ post labels.\footnote{\new{The remaining records could not be retrieved due to missing repositories, deleted records, or unavailable PDS endpoints.}}   
\Cref{tab:label_counts_desc} presents a breakdown of the most frequently applied labels in 2025 along with their counts in our collected dataset. The most prevalent label is \textit{porn} ($7.3$M posts), followed by \textit{sexual} ($2.2$M) and \textit{nudity} ($271$K). Based on these records, we also list distributions of the type of post (root vs. reply), embed-types and languages across the labeled posts in \Cref{tab:label_stats_combined} (Appendix~\ref{Sec: appendix stats other labels}).}

\new{Bluesky posts support five embed types: \texttt{app.bsky.embed.images} attaches up to four images per post; \texttt{app.bsky.embed.video} embeds an uploaded video; \texttt{app.bsky.embed.external} embeds an external URL as a thumbnail; and \texttt{app.bsky.embed.record} embeds a reference to another post, %
optionally combined with media via \texttt{app.bsky.embed.recordWithMedia}. Posts with none of these embed fields are counted as \textit{text-only}.} 

\new{Across all modalities, images dominate across many labels, including \lporn{}, \lsexual{}, \lnudity{} etc., whereas, labels like \lrude{} and \textit{!takedown}, show a high proportion of \textit{text-only} posts. Overall, $88\%$ of labeled posts are root posts, though this varies markedly: labels such as \lporn{}, \lsexual{} and \lnudity{} are overwhelmingly root posts ($91$--$92\%$), whereas posts labeled as \lrude{} ($96\%$), \textit{spam} ($99\%$), and \textit{threat} ($79\%$) skew heavily toward replies.
English is the dominant language across all label categories, ranging from 
$49\%$ (\lselfharm{}) to $97\%$ (\lrude{}) of all posts.}

Our subsequent analyses focus on this set of %
\new{10,681,824} post labels applied by the Bluesky Moderation Service (including country-specific services) covering 13 label values.
We provide the counts of labels on accounts and profiles in Appendix~\ref{Sec: appendix stats other labels}. Similarly, we were able to collect 69 post records, all labeled as \lhide{} by the country-specific BMS. 

\begin{table}[t]
\centering
\caption{BMS label distribution in our collected dataset.
}
\small
\label{tab:label_counts_desc}
\begin{tabular}{ll||ll}
\toprule
\textbf{Label} & \textbf{Count} & \textbf{Label} & \textbf{Count} \\
\midrule
\lporn{}             & 7,360,828 & \lspam{}             &    78,077 \\
\lsexual{}           & 2,243,593 & \lintolerant{}       &    38,021 \\
\lnudity{}           &   271,246 & \lthreat{}           &    11,640 \\
\lrude{}             &   219,712 & \lselfharm{}         &    10,187 \\
\ltakedown{}         &   214,079 & \lhide{}             &     3,520 \\
\lsexualfigurative{} &   134,942 & \lwarn{}             &     2,191 \\
\lgraphicmedia{}     &    93,788 &                      &           \\
\bottomrule
\end{tabular}
\vspace{-4 mm}
\end{table} %

\section{RQ1: Automation and Human Oversight in the Bluesky Moderation Service}
\label{Sec: Time}

\begin{figure*}
    \centering
    \begin{subfigure}[t]{0.3\textwidth}
    \centering
    \includegraphics[width=\linewidth, height=4cm]{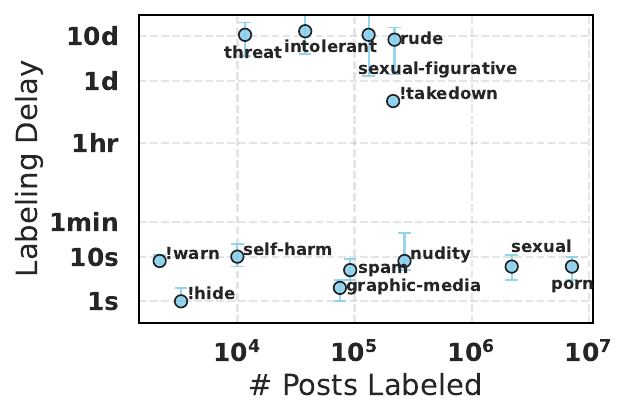}
    \caption{Delay for the BMS labeling}
    \label{fig:delay}
  \end{subfigure}
  \begin{subfigure}[t]{0.3\textwidth}
    \centering
    \includegraphics[width=\linewidth, height=4cm]{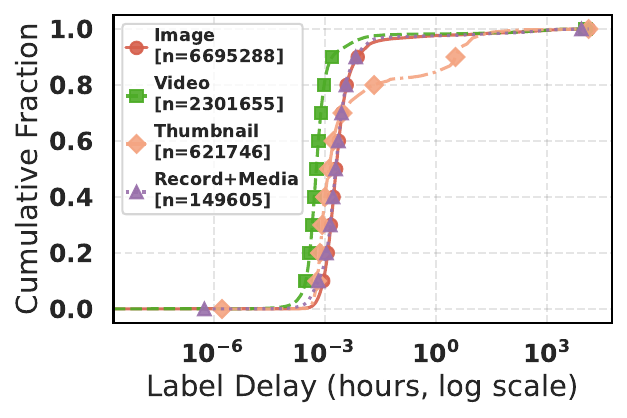}
    \caption{More automation}
    \label{fig:cdf_group2}
  \end{subfigure}
  \begin{subfigure}[t]{0.3\textwidth}
    \centering
    \includegraphics[width=\linewidth, height=4cm]{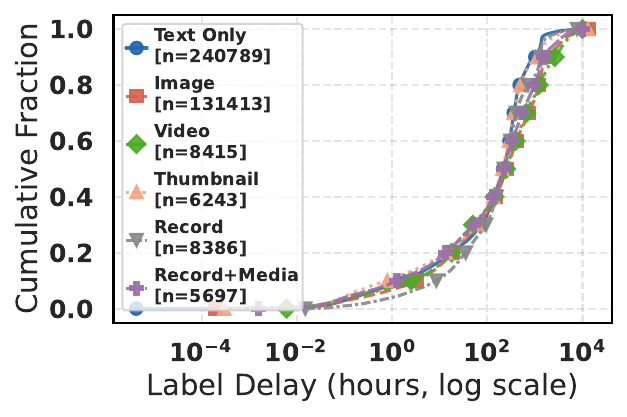}
    \caption{More human oversight}
    \label{fig:cdf_group1}
  \end{subfigure}
  \vspace{-2mm}
  \caption{(a)~Median labeling delay shown with the number of posts labeled. (b)~Cumulative distribution of delays for labels with more automation. (c)~Cumulative distribution of delays for labels with more human oversight. \lselfharm{}, \lporn{}, \lwarn{}, etc. seem to be more automated while \lthreat{}, \lrude{}, \ltakedown{} etc. may involve more human oversight across major post types.}
  \label{fig:delays_combined}
  \vspace{-4mm}
\end{figure*}

\noindent
For any given content moderation system, the Digital Services Act requires platforms to be transparent regarding the degree of human oversight and usage of automated tools in their decision making. 
Bluesky, in its terms of services and transparency reports, clarifies that it uses a human-AI collaborative approach for operationalizing its moderation policies~\cite{Bluesky2025TOS, Bluesky2024Moderation, Bluesky2025Report}. 
However, to this date, apart from transparency reports from platforms, there is no other means to quantify and ascertain the claims made by platforms. 
Therefore, in this section, we analyze the degree of underlying Human-AI collaboration in Bluesky's content moderation pipeline.

\vspace{1mm} \noindent
While the BMS has its label stream (i.e., outcome of its labeling process) transparent and publicly accessible, the exact methodology adopted by the BMS is largely non-transparent and proprietary. 

\noindent
\textbf{Labeling delay as an empirical proxy: }To better understand the degree of human oversight versus automation across different labels, we use labeling delay as an empirical proxy. 
For each labeled post, we define labeling delay as the difference between the time the post was created by the user (\texttt{createdAt}) and the time the labeler applied its label (\texttt{cts}). We hypothesize that:
\begin{quote}
\textit{Labels applied at scale within seconds are likely automated, whereas labels requiring hours or days indicate involvement of significant human oversight.}
\end{quote}

\noindent
Posts were filtered out if they were backdated (created before the cutoff date, i.e., June 2024 or had negative labeling delays due to incorrect \texttt{createdAt} \new{\footnote{Timestamp generally represents the time the record was ``first seen'' by an API server, though it might also be updated if the record is edited. Using only this timestamp would mean that bulk-imported records would all get timestamps very close together around the import time, which may not be correct. \url{https://docs.bsky.app/docs/advanced-guides/timestamps/#createdat}}}
timestamps).

\vspace{1mm} \noindent
\textbf{Automated labels: }For each label, we compute the median labeling delay with inter-quartile ranges. 
\Cref{fig:delay} shows the labeling delay (on Y-axis) along with the number of labeled posts for each label (on X-axis). 
It shows a clear dichotomy in the timeliness of moderation. 
Labels such as \lporn{}, \lsexual{}, \lnudity{}, \lselfharm{}, and \lgraphicmedia{} are applied very quickly, with median delays typically a few seconds. 
For example, the median labeling delay for \lporn{} was $5.5$ s and the upper quartile (Q3) delay was $9.8$ s, already covering more than five million applied labels in our collected dataset.
This pattern indicates that these labels are predominantly applied automatically. 
\Cref{fig:cdf_group2} shows the cumulative distribution of labeling delay for these automated labels for content with different embed types. 
We observe that most of the posts across different embed types: images, video, thumbnails, and embedded records with media are labeled very quickly. 
\new{However, posts with thumbnails exhibit longer delays for a significant fraction of cases as thumbnail embeds refer to external third-party URLs rather than media hosted directly on Bluesky’s CDN, introducing additional fetch and crawl latency before labels can be assigned.}

While the above observations help us identify the potential degree of automation needed in some of the labels, there are some interesting insights if we closely look at the labeling delay distribution. 
\Cref{tab:delays} shows the median, 5th, 25th, 75th, and 95th percentiles of the labeling delay distribution of the different labels. 
For some of the potentially automated labels, e.g., \lnudity{} and \lselfharm{}, the 95th percentile of the labeling delay is of the order of a few days. 
This, coupled with the lower number of these labels, further indicates a greater degree of human oversight on some of these categories, albeit they are predominantly automated as indicated by the median of the distributions. 

\noindent
\textbf{Need for automated moderation: }
Automated labeling is faster and scalable i.e., automated tools can detect harmful posts faster and in larger numbers. 
Faster labeling of harmful content not only reduces their dissemination and exposure to users, but also reduces the mental health and other societal issues for human moderators.
BMS uses automated tools to reduce the spread of sexual and graphic content on its platform to keep it a safer digital place for its users.

\vspace{1mm} \noindent
\textbf{Higher degree of human oversight: }
In contrast, labels such as \lintolerant{}, \lrude{}, \lthreat{}, and \lsexualfigurative{} exhibit much longer median delays, ranging from hours to days, reflecting a significant likelihood of human intervention. 
Furthermore, these labels show larger interquartile ranges, indicating the variability in human moderation times. 
When it comes to variation across content with different multi-media attachments, \Cref{fig:cdf_group1} shows that irrespective of the type of content, the labeling delay follows similar patterns.

\noindent
\textbf{Need for human oversight: }
While moderating harmful content at scale is one of the primary goals of any moderation system, preserving free-speech is also another important aspect of this activity. 
Given the vast array of issues being discussed on social media, some could potentially hurt the sentiments of an individual or a group of people. 
However, often a broader context and perspective are needed to disentangle absolute violations of community guidelines. 
For example, criticism of political parties or opinions should be allowed to preserve and empower free-speech. 
However, uncivil, impolite, and abusive posts, even in such contexts, need to be labeled (if not removed). 
At the same time, the delay we observe for some of the automated labels could be because of the inherent subjectivity of some of the situations. 
For example, although posts detailing war-time atrocities~\cite{gillespie2018custodians} or breastfeeding~\cite{Sweney2008Mums, Arthur2012Facebook} could be identified as \lgraphicmedia{} or \lnudity{} respectively, 
whether they \textit{should} be labeled is often subjective.
Our analyses bring to light the thoughtful effort taken up by the BMS for detecting harms prevalent on Bluesky. 
\\
\new{Additional analyses on country-specific moderation services and Bluesky's label redressal mechanism, both of which exhibit higher human oversight, are provided in Appendix~\ref{Sec: appendix delays}.}

\noindent
\textbf{Validation of our approach: }
\new{
While labeling delay serves as a proxy for distinguishing automated from human-in-the-loop labels, AT Protocol offers no independent means to validate this. Nevertheless, Bluesky's 2025 transparency report~\cite{Bluesky2025Report} corroborates our findings: \lrude{}, \lintolerant{}, and \lthreat{} are confirmed as manually applied, while \lnudity{}, \lporn{}, \lsexual{}, \lgraphicmedia{}, and \lselfharm{} have significantly lower human oversight. Their reported manual-application rates for \lselfharm{} and \lnudity{} (8\% and 18\%, respectively) align with our observation that delays for these labels extend to hours or days. For \lsexualfigurative{}, we observe high delays suggesting human oversight despite the 2025 report listing 0\% manual application~\cite{Bluesky2025Report}. Their 2024 moderation report~\cite{Bluesky2024Moderation} resolves this discrepancy, clarifying that this label was predominantly applied when users appealed automated misclassifications of figurative art, confirming that the observed delays do reflect human involvement.
}

\begin{table}
\centering
\caption{Labeling delay for different labels.}
\small
\setlength{\tabcolsep}{3pt}
\renewcommand{\arraystretch}{0.6}
\resizebox{\linewidth}{!}{%
\begin{tabular}{p{3cm}rrrrr}
\toprule
\textbf{Label (N)} & \textbf{5\%ile} & \textbf{25\%ile} & \textbf{Median} & \textbf{75\%ile} & \textbf{95\%ile} \\
\midrule
\midrule
\multicolumn{6}{c}{\textbf{Potentially automated label}}\\
\midrule
\midrule
\textbf{porn} (7,198,690) & 1.3 s & 2.8 s & 5.5 s & 9.8 s & 55.0 s \\
\midrule
\textbf{sexual} (2,200,976) & 1.5 s & 3.3 s & 6.2 s & 11.1 s & 2.7 min \\
\midrule
\textbf{nudity} (267,334) & 2.1 s & 4.6 s & 8.1 s & 33.9 s & 32.0 d \\
\midrule
\textbf{graphic-media} (91,832) & 1.5 s & 3.0 s & 4.9 s & 9.1 s & 52.4 min \\
\midrule
\textbf{self-harm} (9,972) & 2.1 s & 5.7 s & 10.2 s & 19.0 s & 10.3 d \\
\midrule
\textbf{spam} (74,984) & 0.7 s & 1.1 s & 1.7 s & 2.9 s & 12.2 s \\
\midrule
\textbf{!warn} (2,164) & 5.8 s & 7.0 s & 8.1 s & 10.8 s & 26.8 s \\
\midrule
\textbf{!hide} (3,294) & 0.5 s & 0.9 s & 1.3 s & 2.5 s & 15.7 s \\
\midrule
\midrule
\multicolumn{6}{c}{\textbf{Labels with higher human oversight}}\\
\midrule
\midrule
\textbf{!takedown} (213,677) & 2.4 hr & 7.3 hr & 8.5 hr & 9.7 hr & 13.8 d \\
\midrule
\textbf{sexual-figurative} (131,891) & 20.4 min & 1.3 d & 10.8 d & 43.1 d & 150.7 d \\
\midrule
\textbf{intolerant} (37,894) & 2.5 hr & 4.0 d & 13.0 d & 47.0 d & 80.4 d \\
\midrule
\textbf{rude} (219,557) & 11.7 min & 1.4 d & 8.4 d & 15.7 d & 52.3 d \\
\midrule
\textbf{threat} (11,601) & 2.0 hr & 3.6 d & 10.8 d & 20.5 d & 58.7 d \\
\midrule
\bottomrule
\end{tabular}}
\label{tab:delays}
\vspace{-4mm}
\end{table}

\section{RQ2: Evaluation of Effectiveness of the Bluesky Moderation Service}\label{Sec: Accuracy}
The label delay analysis establishes that the BMS operates as two distinct pipelines: an automated system handling visual content and a human oversight system handling specific labels. 
A natural question follows: \textit{how accurate (or effective) is each pipeline and the BMS labeling of harmful content overall?} 
To answer this, we evaluate the system along two dimensions -- precision (of what the BMS flags, how much is actually harmful?) and recall (of all harmful content, how much does the BMS flag?)%

\subsection{Precision and Recall of the BMS}

\noindent\textbf{Evaluation sets and annotation procedure: } 
Measuring precision requires a sample drawn exclusively from contents flagged by the BMS, so that human judgments can verify their accuracy. 
On the other hand, measuring recall requires a random sample drawn from the content stream i.e the firehose, so that humans can flag unsafe or harmful contents and compare the same with what the BMS might have flagged in the random sample.  
To evaluate precision, we randomly sample 1,000 labeled posts from our dataset (\textbf{labeled set}), stratified across all nine harm categories (\lintolerant{}, \lrude{}, \lthreat{}, \lselfharm{}, \lgraphicmedia{}, \lporn{}, \lsexual{}, \lnudity{}, \lsexualfigurative{}) to ensure coverage of both automated and human-oversight labels. 
Furthermore, to evaluate recall, we randomly sample 1,000 posts from the Bluesky firehose (input of the labeler) data in 2025 (\textbf{random set}).

\noindent\textbf{Annotation procedure: } 
Given the %
BMS policies, two coauthors marked the posts in both the %
evaluation sets as safe and unsafe. 
If they mark a post to be unsafe, they were additionally asked which harm label should be assigned to the given post and why. 
In case of disagreement, %
a third co-author performed the same task acting as a tiebreaker. 
Across the full evaluation set (n=2,000), the primary annotators achieved an inter-annotator agreement, cohen's kappa $\kappa = 0.802$ in the binary task of determining whether a post is safe or unsafe. 
They agreed on 1,827 (i.e., 91.35\%) of the posts and the tiebreaker annotator annotated the remaining posts.

\begin{figure}[t]
    \centering
    \includegraphics[width=0.7\columnwidth, height=4cm]{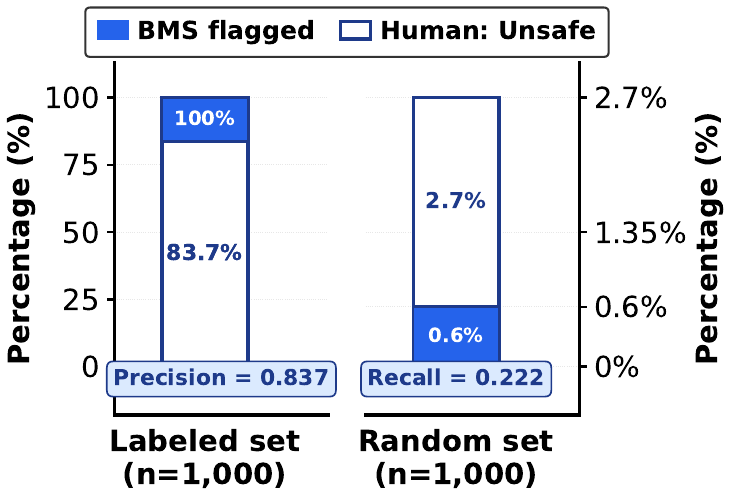}
    \vspace{-4mm}
    \caption{Unsafe content in the labeled set (left) and random firehose set (right). Blue = Flagged by the BMS, white = adjudicated unsafe by majority human annotators. Bluesky achieves high precision (0.837) on the labeled set, but low recall (0.222) on the random firehose. %
    }
    \label{fig:precision_recall}
    \vspace{-4mm}
\end{figure}

\noindent\textbf{Precision of the BMS:} Table~\ref{tab:precision_label} shows the detailed results on the evaluation set that contained 1000 posts labeled by the BMS. 
Note that we report the agreement only for the binary setting i.e., agreement between our majority voting for a post (safe vs. unsafe) and if the BMS has labeled it (irrespective of the label). 
Across all label categories, %
annotators agreed with the BMS' assessment in 83.7\% posts, indicating a significant percentage of posts flagged by the BMS to be genuinely unsafe i.e., the BMS has high precision (0.837) in detecting harms on Bluesky. 
But, this aggregate conceals substantial variation across labels and post structures. %

\begin{table}[t]
\centering
\caption{BMS' precision per label broken down by post type.}
\label{tab:precision_label}
\scriptsize
\setlength{\tabcolsep}{3pt}
\renewcommand{\arraystretch}{0.8}
\resizebox{\columnwidth}{!}{
\begin{tabular}{l rr rr rr}
\toprule
 & \multicolumn{2}{c}{\textbf{Total}} & \multicolumn{2}{c}{\textbf{Root}} & \multicolumn{2}{c}{\textbf{Reply}} \\
\cmidrule(lr){2-3} \cmidrule(lr){4-5} \cmidrule(lr){6-7}
\textbf{Label} & \textbf{n} & \textbf{Precision} & \textbf{n} & \textbf{Precision} & \textbf{n} & \textbf{Precision} \\
\midrule
\multicolumn{7}{l}{\textbf{Automated Labels}} \\
\hspace{4pt}\lporn{}          & 111 & 1.000 & 102 & 1.000 &   9 & 1.000 \\
\hspace{4pt}\lsexual{}        & 111 & 0.856 &  95 & 0.884 &  16 & 0.688 \\
\hspace{4pt}\lnudity{}        & 111 & 0.937 & 100 & 0.940 &  11 & 0.909 \\
\hspace{4pt}\lselfharm{}      & 111 & 0.919 &  99 & 0.939 &  12 & 0.750 \\
\hspace{4pt}\lgraphicmedia{}  & 111 & 0.802 &  86 & 0.814 &  25 & 0.760 \\
\cmidrule(lr){2-7}
\hspace{4pt}\textbf{Total}    & \textbf{555} & \textbf{0.903} & \textbf{482} & \textbf{0.919} & \textbf{73} & \textbf{0.795} \\
\midrule
\multicolumn{7}{l}{\textbf{Manual Labels}} \\
\hspace{4pt}\lintolerant{}         & 111 & 0.667 &  23 & 0.609 &  88 & 0.682 \\
\hspace{4pt}\lrude{}               & 112 & 0.509 &   3 & 0.667 & 109 & 0.505 \\
\hspace{4pt}\lthreat{}             & 111 & 0.892 &  16 & 0.812 &  95 & 0.905 \\
\hspace{4pt}\lsexualfigurative{}   & 111 & 0.955 & 104 & 0.952 &   7 & 1.000 \\
\cmidrule(lr){2-7}
\hspace{4pt}\textbf{Total}         & \textbf{445} & \textbf{0.755} & \textbf{146} & \textbf{0.877} & \textbf{299} & \textbf{0.696} \\
\midrule
\textit{Overall} & \textbf{1000} & \textbf{0.837} & \textbf{628} & \textbf{0.909} & \textbf{372} & \textbf{0.715} \\
\bottomrule
\end{tabular}}
\end{table}

In the automated labels, our binary annotation (safe vs unsafe) and the BMS' flagging achieve an agreement of 90.3\% (i.e., BMS' precision of 0.903). %
In the \lporn{} label, the BMS achieves a perfect precision (1.000, n=111): every post the BMS flagged was annotated as unsafe by majority of the annotators. 
Similarly, in \lselfharm{} (0.919), \lnudity{} (0.937), and \lsexual{} (0.856), the BMS achieves high precision in detecting harmful content that is inherently sexual or self-harm related. 
However, its precision drops to 0.802 for \lgraphicmedia{}. 
On the other hand, the BMS' precision among manual labels drops to 0.755. 
Among manual labels, the BMS achieves high precision in \lsexualfigurative{} (0.955), and \lthreat{} (0.892). 
However, its precision drops significantly for \lintolerant{} (0.667) and \lrude{} (0.509).

\noindent\textbf{Root posts vs. Replies: }
In further analyses, we observed that a large percentage of the posts labeled by us in \lintolerant{} (79\%), \lrude{} (97\%), and \lthreat{} (85\%) are in fact replies to some other posts. 
This indicates a potential lack of additional context which Bluesky's moderators might have had (e.g., the root post and/or authors to which these replies were directed) which was unfortunately not provided to our annotators as they annotated each individual post independently treating them as an isolated post. In fact, in many such cases, we observe that annotators have clearly marked that they need more context to adjudicate such posts as unsafe. 
Overall, while our annotations agree with the BMS's flagging in 71.5\% cases for posts which were replies, the agreement increases to 90.9\% for root posts.

\noindent\textbf{Disagreement analysis:%
} 
To characterize the false positives qualitatively, we examined a random sample of 50 posts judged \textit{safe} by majority vote despite carrying a Bluesky label and observe some noticeable patterns. 
Firstly, as discussed earlier \textit{subjectivity and lack of context}-- posts where annotators evaluated reply text without access to the parent thread, making it impossible to assess tone, intent, or conversational register -- content that could or could not be harmful depending on the conversational context in which it is produced. 
Secondly, \textit{ambiguous visual content}-- images that may resemble harmful content, but carry benign intent in context e.g., body art mimicking self-harm imagery, film and media stills flagged as graphic, pop album covers without sexual intent etc. for a notable share of automated label false positives. 
Overall, these cases also reflect inherent subjectivity of the labeling task: annotators disagreed substantially on what constitutes harmful expression for labels such as \lrude{} and \lintolerant{}, underscoring that such judgments are sensitive to individual thresholds and cultural background. 

\noindent\textbf{Recall of the BMS: } 
In the \textbf{random set}, we label a total of 27 posts to be unsafe i.e., 2.7\% of all posts in the random set of 1000 posts. 
Out of the 27, only 6 had been labeled by the BMS,  
yielding a recall of 0.222. 
On the positive side, the BMS achieved a 100\% precision in the random set where all posts labeled by the BMS are adjudicated as unsafe by human annotators. 
Notably, all 6 posts (flagged by the BMS) carried automated labels. 
Therefore, %
its recall for manually applied labels for this random set is 0. 
This is consistent with the operational reality of human-oversight labels: moderators usually only review reported or surfaced content and cannot read every post at a platform scale, so content of this type can go unlabeled unless explicitly reported. 
Of the 10 posts human annotators identified as automated-label harm, Bluesky caught 6 and missed 4, resulting in a recall of 0.600. 
Such low recall for automated labels
raises a question: %
\textit{why does Bluesky's automated pipeline fail to catch these posts?} 

\subsection{Blindspots of Automated Moderation}
\label{sec:blindspots}
To diagnose the potential reasons for low recall of the BMS' automated labeling, %
we first describe the technical architecture underlying Bluesky's automated moderation, then design a setup to study the potential blindspot of the pipeline. 

\vspace{1mm} \noindent
Bluesky's automated moderation pipeline 
is powered by two components: %
\textbf{(i) Hive AI}\footnote{\url{https://docs.thehive.ai/docs/visual-content-moderation}}, a commercial API providing multi-head vision classifiers for moderation. 
When a post containing an image (individual frames in case of a video) is submitted, Bluesky calls Hive API %
to get a set of class-score pairs for the image, with each class representing a specific visual concept (e.g., \texttt{yes\_sexual\_activity}, \texttt{yes\_self\_harm}, \texttt{very\_bloody}) with a confidence score between 0 and 1. Note that 
Hive does not return a single label; it returns a probability distribution across a hundred %
visual categories simultaneously. 
\textbf{(ii) Automod}\footnote{\url{https://github.com/bluesky-social/indigo/blob/main/automod/visual/hiveai_client.go}}, Bluesky's open-source rule engine that translates Hive's raw scores into Bluesky's label vocabulary via hard-coded rules\footnote{The Automod code was open-sourced in April 2024. It is difficult to establish whether the BMS still uses the exact rules or not. However, this is the best available approximation for the automatic moderation rules provided by Bluesky itself.}.
For example, \texttt{yes\_self\_harm} $\ge$ 0.96 triggers the \lselfharm{} label, while sexual content follows a priority cascade: \lporn{} is checked (using multiple head scores) before \lsexual{}, which is checked before \lnudity{}. 
Critically, Automod reads only a subset of Hive’s output heads; scores on all others are ignored, regardless of magnitude. Details %
on Hive classes and Automod are in Appendix~\ref{Sec: appendix hive}.

\vspace{1mm} \noindent
\textbf{Experimental setup:}
To investigate the low recall for automated labels applied by the BMS, we start with our labeled set of 1,000 posts. 
We filtered to posts carrying one of the five automated labels and containing a single image, yielding a set of 336 labeled posts distributed across \lporn{} (78), \lsexual{} (67), \lnudity{} (72), \lselfharm{} (57), and \lgraphicmedia{} (62). 
For each labeled post, we retrieved the most semantically similar image post from a firehose set carrying no Bluesky label, yielding a second set of 336 unlabeled posts. 
This firehose set is created by randomly sampling 40M posts between March and June 2025. 
Similarity is computed using cosine similarity between embeddings of labeled post and firehose posts (indexed using faiss) generated using \texttt{Qwen3-VL-Embedding-2B}. Details are provided in Appendix~\ref{Sec: appendix firehose}.
These unlabeled posts were selected for their visual and semantic proximity to known harmful content, making them very likely candidates for missed moderation, thus helping us investigate recall failures. 
Both sets were passed through Bluesky's Automod pipeline via the Hive API\footnote{We investigate on a smaller set since Hive AI API has a rate limit of 100 requests/day for developers.}, obtaining raw classifier scores across all heads.

\noindent\textbf{Validating Automod as a proxy:} Before analyzing the unlabeled set, we verified that Automod faithfully reproduces Bluesky's labeling decisions. 
Of the 336 labeled posts, Automod independently flagged 311 (92.6\%), confirming high agreement with Bluesky's deployed moderation system. 
This slight mismatch could be due to API versions: developers can only access Hive's V3 API, whereas enterprises have access to V2. Nonetheless, our findings show that we can treat Automod as a reliable proxy for Bluesky's automated pipeline in subsequent analyses.

\noindent\textbf{Identifying Missed Content:} Of the 336 most similar unlabeled posts, Automod flagged only 47 (14.0\%), leaving 289 posts flagged by neither BMS nor Automod. We manually annotated these 289 posts: two annotators independently judged 36 of them as unsafe content ($\kappa$=0.63, substantial) Bluesky's system had missed entirely. This corresponds to 10.7\%\footnote{The relatively low proportion of unsafe posts, even among the closest semantic matches to known harmful content is expected: semantic similarity is computed over a general embedding space where proximity reflects similarity across all topical and visual features jointly, and not exclusively along the harm subspace.} of all 336 similar unlabeled posts, establishing a lower bound on recall loss for automated labels. 

\begin{figure}[t]
    \begin{subfigure}[t]{0.48\columnwidth}
        \centering
        \includegraphics[width=\linewidth, height=3.5cm]{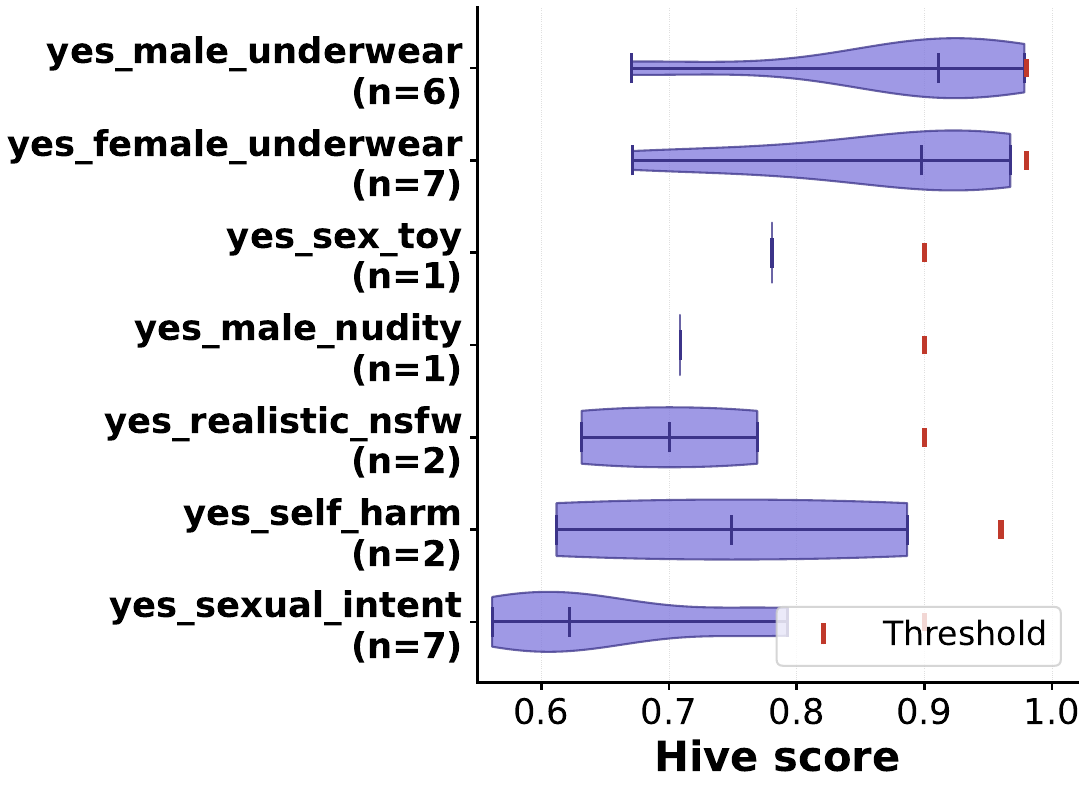}
        \caption{H1: close threshold misses}
        \label{fig:h1_violin}
    \end{subfigure}
    \hfill
    \begin{subfigure}[t]{0.48\columnwidth}
        \centering
        \includegraphics[width=\linewidth, height=3.5cm]{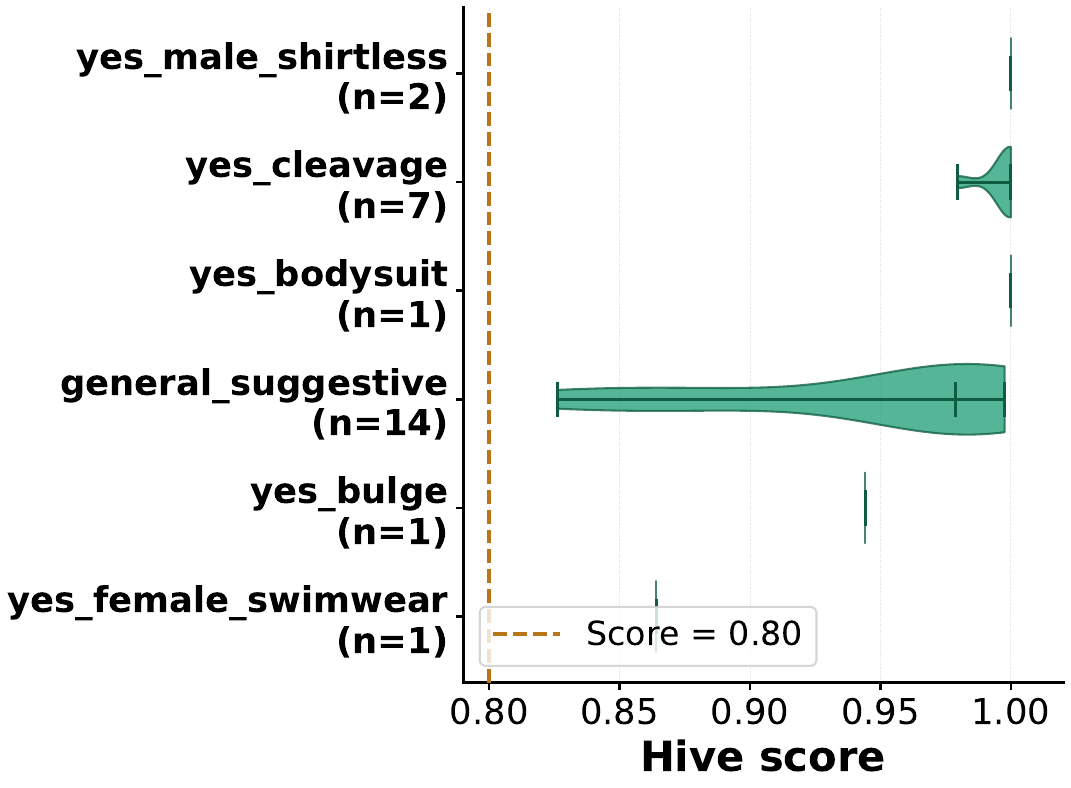}
        \caption{H2: Rule set gap}
        \label{fig:h2_violin}
    \end{subfigure}
    \vspace{-2mm}
    \caption{%
    (a) Hive scores on Automod-mapped heads sorted by median gap to hard-coded thresholds (red markers); some harmful posts narrowly miss.
    (b) Hive scores on heads outside Automod's rule set sorted by median score; Automod does not act despite near-perfect scores.
    }
    \label{fig:blindspots}
    \vspace{-4mm}
\end{figure}

\noindent\textbf{Blindspots:} We examined the 36 missed posts against Hive's classifier outputs and identified two failure modes:

\noindent\textit{H1 -- Close threshold misses} (22 posts, 61.1\%): Hive's visual concept classes %
produce confidence scores in the near-miss zone [0.55, threshold) -- the classifier detects the content as potentially harmful on heads Automod uses, but falls short of threshold i.e it is excluded due to a conservative threshold. 
Figure~\ref{fig:blindspots} (left) shows the score distributions for each near-miss head, sorted by median gap to threshold. 
Heads like \texttt{yes\_sexual\_intent} and \texttt{yes\_self\_harm} have median gaps of 0.28 and 0.21, while \texttt{yes\_female\_underwear} (median gap=0.08) and \texttt{yes\_male\_underwear} (median gap=0.07) are the closest -- consistently below Automod's threshold(0.98). %

\noindent\textit{H2 -- Rule set gap} (14 posts, 38.9\%): No visual concept class from Hive (considered by the Automod) approaches the threshold, but classes outside Automod's rule set score $\geq$ 0.80. 
Figure~\ref{fig:blindspots} (right) shows these soft classes sorted by median score. 
\texttt{yes\_cleavage} (n=7, median=1.000) and \texttt{yes\_male\_shirtless} (n=2, median=1.000) achieve perfect or near-perfect scores, while \texttt{general\_suggestive} dominates by volume (n=14, median=0.979). 
The content passes through Automod entirely because no rule acts on the presence of these visual categories.

\noindent\textbf{Brittleness of rule-based moderation: }
These findings expose the brittleness of a rule-based system sitting on top of classifiers. Automod's rules are hard-coded: with a fixed (and potentially arbitrary) threshold and fixed set of heads. 
This rigidity could be by design: having explicit rules is auditable, explainable, and deployable at scale. 
However, a post scoring 0.89 on \texttt{yes\_sexual\_activity} is treated identically to one scoring 0.10, despite being qualitatively far closer to a flagged post (H1). 
A post with strong signal on, say, \texttt{yes\_bulge} or \texttt{general\_suggestive} is missed by such a system not because the content is ambiguous, but because there is no rule written for those heads (H2). 
In both cases, the failures are due to the translation layer between continuous scores and binary moderation decisions. \neww{In this regard, Vision-Language models (VLMs) offer a more general approach to moderation: by conditioning on the policy text itself, they can reason toward its intent rather than being confined to a fixed set of thresholds and heads \cite{majumdar2026moderation}.}

\color{black}
\section{RQ3: Landscape of Harms Detected by the Bluesky Moderation Service}
\label{Sec: Transparent}

\begin{figure*}[t]
  \centering
  \begin{subfigure}[t]{0.33\textwidth}
    \centering
    \includegraphics[width=\linewidth, height = 3.5cm]{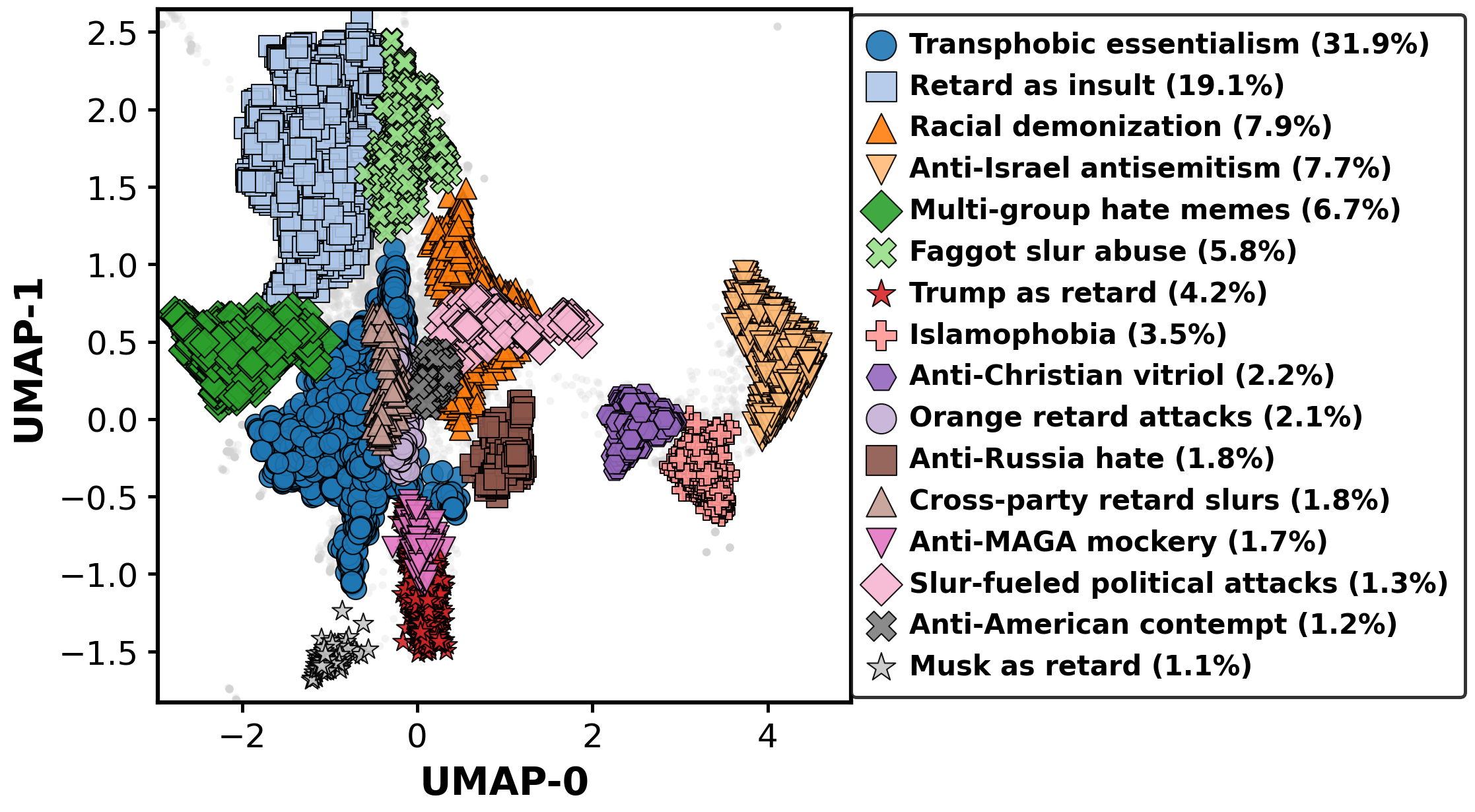}
    \caption{Intolerant}
    \label{fig:intolerantA}
  \end{subfigure}
  \hfill
  \begin{subfigure}[t]{0.33\textwidth}
    \centering
    \includegraphics[width=\linewidth, height = 3.5cm]{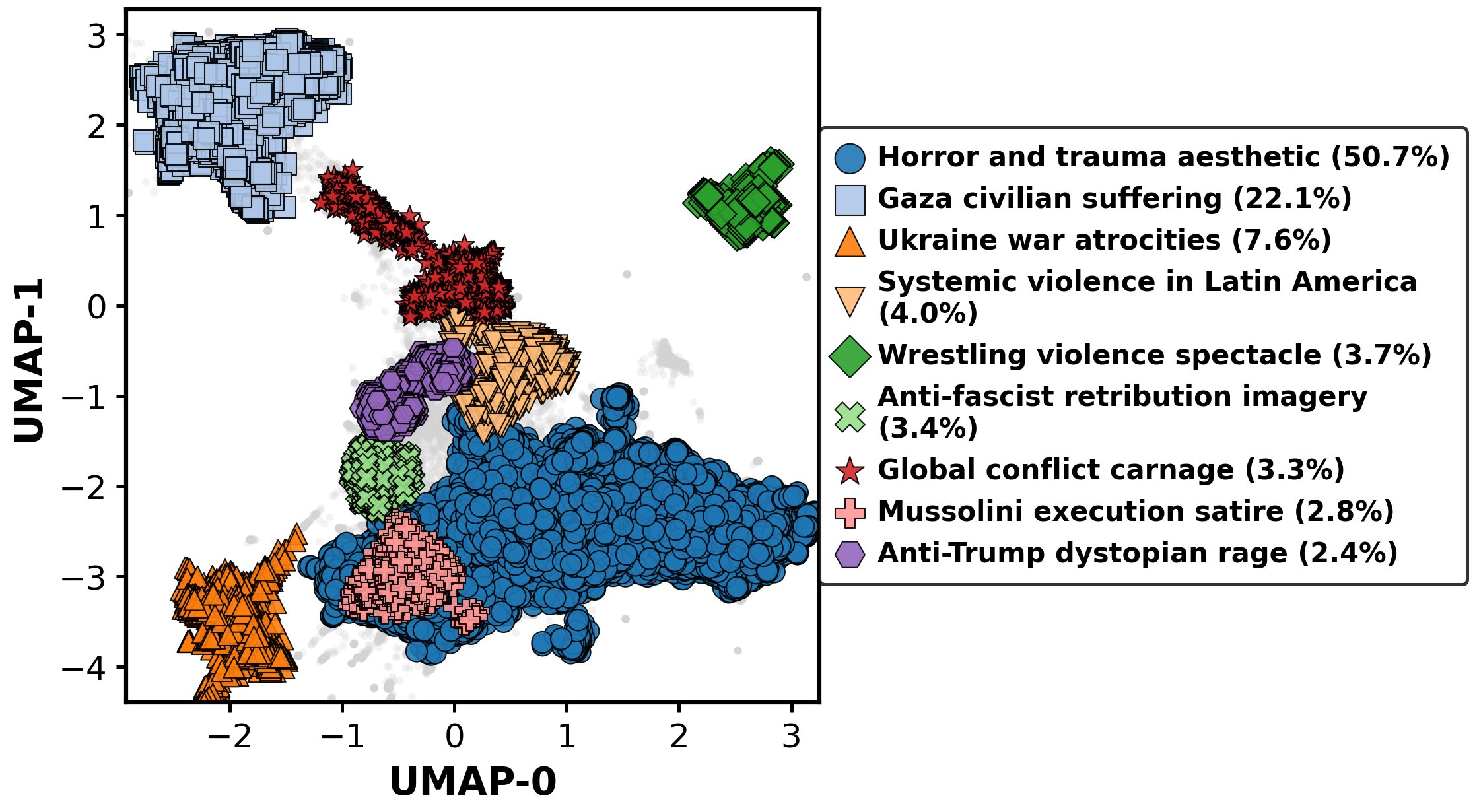}
    \caption{Graphic media}
    \label{fig:graphicmediaB}
  \end{subfigure}
  \hfill
 \begin{subfigure}[t]{0.33\textwidth}
    \centering
    \includegraphics[width=\linewidth, height = 3.5cm]{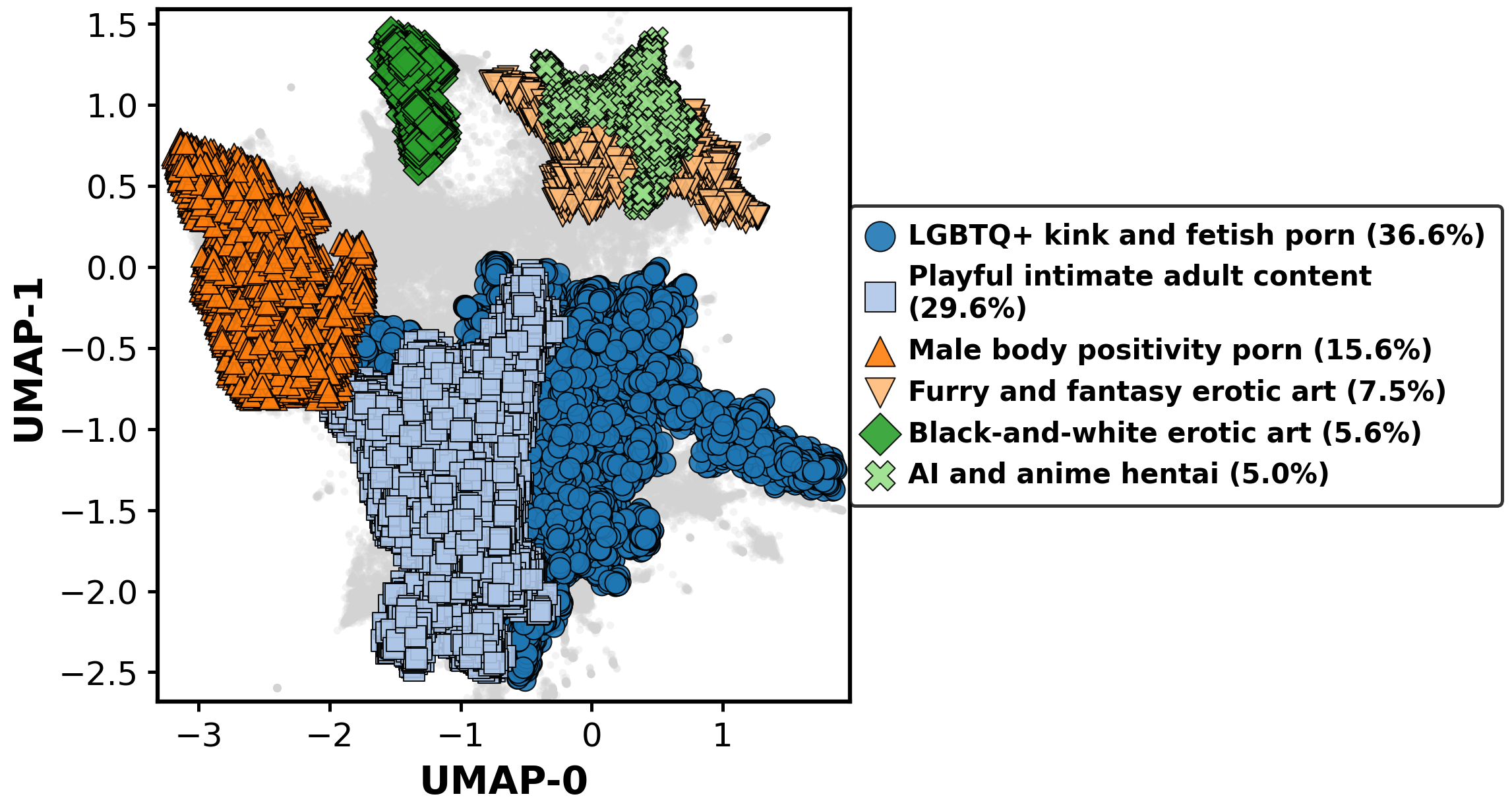}
    \caption{Porn}
    \label{fig:pornC}
  \end{subfigure}
    \vspace{-2 mm}
  \caption{Visualizations of prominent clusters for representative labels that detect (a)~Harms to protected groups, individuals, and civic discourse (b)~Harms to physical and mental well-being, (c)~Harms to minors from sexual content. %
  }
  \label{fig:three_umap_plots}
  \vspace{-4 mm}
\end{figure*}

\noindent
The BMS provides each label along with its policy description (see Table~\ref{tab:all_label_values}). 
As described in the previous section, these descriptions are often understandably abstract or insufficiently detailed, 
raising the difficulty of understanding any additional intended nuances. 
However, due to the availability of the label stream, the outcome of the labeling process is transparent. 
Such transparency not only enables better understanding of how the mentioned policy descriptions are interpreted and operationalized by moderators, but also empowers us to have a better understanding of the kind of harms prevalent on Bluesky. 
In this section, we aim to analyze the posts labeled by the BMS to characterize its detected harms.

\subsection{Clustering of Labeled Posts}

To analyze patterns in posts labeled by the BMS, we perform a multi-modal, unsupervised semantic analysis of the labeled content using clustering. 
Our goal is to examine the dominant content themes within each label to understand the prevalent online harms and compare these themes with the label's policy description to understand the potential interpretations being operationalized. 

\noindent\new{\textbf{Data preparation: }For each label, we retain posts that carry text, images, or videos. We exclude\footnote{We observed a lot of duplicates for \ltakedown{} which contained 210,516 posts but only 31,129 unique records after deduplication.}  posts where records were available, but failed downloading blobs or with the embed type  \texttt{app.bsky.embed.record} (i.e., pure quote-posts or record embeds with no standalone media), which have incomplete information. 
The final post counts used for clustering per label are reported in \Cref{tab:hyperparam} in the Appendix.}

\noindent\new{\textbf{Embeddings and clustering: }
We generate a 2048-dimensional embedding for each post using \texttt{Qwen3-VL-Embedding-2B}~\cite{li2026qwen3}, a vision-language model that jointly encodes text and media (images, video frames (R3-6)). %
These embeddings are projected into a low-dimensional manifold using UMAP~\cite{mcinnes2020umapuniformmanifoldapproximation} with cosine distance, then clustered using HDBSCAN~\cite{campello2013density}. Hyperparameters (\textit{n\_neighbors}, \textit{min\_cluster\_size}, \textit{min\_samples}) are jointly tuned per label via Bayesian optimization (Optuna TPE sampler~\cite{10.1145/3292500.3330701}), evaluated using the Density-Based Clustering Validation (DBCV) score~\cite{doi:10.1137/1.9781611973440.96}, an intrinsic measure rewarding internally dense, well-separated clusters. DBCV scores range from $-1$ (poor) to $+1$ (ideal), and the configuration maximizing this score per label is selected. For labels exceeding one million posts (\lporn{} and \lsexual{}), we uniformly sample 300,000 posts and perform all steps %
on them, ensuring computational efficiency while preserving semantic diversity for meaningful clusters.}

\new{Full details of the hyperparameter tuning and selected hyperparameters are in Appendix~\ref{Sec: appendix clusters}. Posts not assigned to any cluster by HDBSCAN are designated as noise; these represent posts that do not belong to any sufficiently dense region of the embedding space and are excluded from our analysis. Noise fractions vary substantially across labels (\Cref{tab:hyperparam}), reflecting heterogeneity in some label categories.}

\noindent\new{\textbf{Cluster interpretation.}
After obtaining the resulting clusters, we analyze the posts within each cluster. Due to the sensitivity of labeled content, particularly label values such as \lgraphicmedia{} and \lselfharm{}, we adopt an automated pipeline to avoid exposing human annotators to potentially harmful material~\cite{arsht2018human, spence2023psychological} and validate it on a subset of clusters. Posts are sampled per cluster using the Cochran~\cite{cochran1977sampling} finite-population-corrected formula (Eq.~\ref{eq:cochran} in the Appendix~\ref{Sec: appendix clusters}), weighted by HDBSCAN soft membership probabilities, and passed to \texttt{Qwen3-VL-32B-Instruct} using a map-reduce summarization strategy to generate a cluster description that describes the theme of posts in that cluster which is then used to get a name for the cluster. The full pipeline including the sampling, prompts used, LLM generation parameters is described in Appendix~\ref{Sec: appendix clusters}.}

\noindent\new{\textbf{Human validation.}
For human validation of the assigned cluster names, we randomly selected 31 clusters across 7 labels and asked three annotators %
to independently assess the appropriateness of the cluster names. 
Each annotator was provided with 10 sample posts from the cluster along with the cluster name and was asked to rate the cluster names
on a five-point Likert scale ranging from \textit{very inappropriate} to \textit{very appropriate}. 
Cluster names received a mean rating of $\mu = 4.58$ (SD\,$= 0.63$), with 92.5\% of
ratings at \textit{appropriate} or \textit{very appropriate} ($\geq 4$) with substantial inter-annotator agreement (Krippendorff's $\alpha = 0.686$)~\cite{landis1977measurement} indicating our pipeline produces cluster names largely faithful to
underlying post content. Full details are in
Appendix~\ref{Sec: appendix clusters}.}

\subsection{Detected Harmful Contents by Bluesky}

\Cref{fig:three_umap_plots} shows the top clusters observed in \lintolerant{}, \lgraphicmedia{} and \lporn{} labels, respectively. %
(Please refer to Figure~\ref{fig:six_umap_plots} in Appendix~\ref{Sec: appendix clusters} for other labels). 
Next, we describe the prominent topics within each moderation label. %

\vspace{1 mm}
\noindent
\textbf{Harms to protected groups, individuals, and civic discourse: }
\Cref{fig:intolerantA} shows the prominent clusters found among posts labeled as \lintolerant{}. 
The BMS defines intolerance as `discrimination against protected groups'. 
To this end, our clusters show potential interpretations and operationalization of this abstract nuance. 
For example, we observe clusters of posts targeting people based on their gender identities (cluster: Transphobic essentialism) e.g., LGBTQ+ identities (cluster: faggot slur abuse), religious identity (cluster: islamophobia, anti-christian vitriol) e.g., anti-semitic, anti-muslim content, racial identity (cluster: racial demonization) e.g., toward black, white and immigrant communities, political preference targeting political figures and their supporters, with posts centered on mocking Trump (clusters: Trump as retard, Orange retard attacks) and MAGA supporters (cluster: Anti-MAGA mockery), alongside broader cross-partisan attacks (clusters: Cross-party retard slurs, Slur-fueled political attacks).%

\lrude{} and \lthreat{} labels are an interesting case study, which could be thought of as a stronger degree of incivility than intolerance. 
Clusters in the \lrude{} label include posts characterized by personal attacks, vulgar insults, often reflecting deep political polarization. \new{These manifest across several rhetorical patterns: targeting individuals with slurs and commands (clusters: Eat s**t insults, Go f*** yourself, F*** off demands), \textit{identity-based attacks} deploying misogynistic slurs or accusations of pedophilia and nazism to dehumanize opponents (clusters: Misogynistic slurs, Pedophilia accusations, Nazi/fascist slurs), and \textit{politically charged hostility} directed at figures and movements across the spectrum (clusters: Anti-MAGA vitriol, Political tribal rage). 
Other clusters reflect outrage over ongoing geopolitical conflicts (clusters: Israel-Palestine hate, Genocide accusations), generalized contempt expressed through mockery (clusters: Idiot accusations, Stupidity mockery, Loser attacks).}

Clusters within \lthreat{} label include similar topics, with an even higher degree of incivility wherein authors express extreme hostility and violent ideas, including calls for death and/or violence, celebrating the pain of others, directed at political figures (clusters: Anti-Trump death wishes, Broad political murder fantasies, Cross-party political violence), geopolitical adversaries (clusters: Anti-Putin nuclear rage, Anti-Israel genocidal rhetoric), institutions (cluster: Anti-ICE assassination calls), and public figures such as billionaire elites (cluster: Anti-Musk vigilante memes).

\vspace{1 mm}
\noindent
\textbf{Harms to physical and mental well-being: } 
For \lgraphicmedia{} label (shown in Figure~\ref{fig:graphicmediaB}), clusters group posts containing explicitly violent, gory imagery, often documenting real-world atrocities or leveraging graphic visuals to express extreme political sentiment. 
Some clusters revolve around historical execution photography repurposed for contemporary political hostility; others are about the war-related civilian deaths in Gaza or Ukraine, where graphic visuals of injured or deceased people (including children) are shown. 
Several clusters also include graphic stills from horror films or wrestling, showing blood or stylized gore, mixing violence with dark humor. We also observe self-harm-related posts getting labeled as \lgraphicmedia{}, especially posts with images of self-inflicted wounds. 

Clusters within the related \lselfharm{} label include posts explicitly depicting or discussing self-inflicted injury, especially cutting, often including hashtags such as \#sh or \#shtwt. Other categories of posts include dark-humor memes involving suicidal themes, such as repeated edits of a man holding a gun to his head juxtaposed with cartoon characters in meme format. 
\neww{In contrast, one cluster includes posts sharing the historical 1863 photograph of an enslaved man’s scarred back as political commentary. These posts are grouped under \lselfharm{} despite using the image to protest the erasure of slavery history (cluster: Slavery scars vs.\ Trump erasure). Since \lselfharm{} is a largely automated label, this case suggests that automated moderation systems can misclassify the \textit{type of harm} depicted in content based on visual appearance alone, without accounting for the broader context or intent behind the imagery.}

\vspace{1 mm}
\noindent
\textbf{Harms to minors from sexual content: }
The BMS uses four different labels to identify content that may be sexually explicit. 
\Cref{fig:pornC} shows the prominent clusters of \lporn{} label. 
Contents within these clusters include highly graphic content spanning kink and fetish pornography, explicit male body-focused content, AI-generated and anime-style hentai, and furry or fantasy erotic art with exaggerated anatomy—frequently accompanied by monetization links to platforms such as OnlyFans and Fansly. 
The \lporn{} label detects overtly explicit and commercial adult material on Bluesky and is the top applied label.
On the other hand, the \lsexual{} label detects content which could be deemed inappropriate but is often significantly less explicit. 
These clusters contain labeled posts encompassing erotic body-focused self-presentation, monochrome sensual art, anime fan art, and queer NSFW art communities, with both real-user imagery and AI-generated content, which may pose difficult challenges to content moderation systems.

Similarly, \lnudity{} label detects a broad spectrum of non-explicit nude content, from fine art nude photography, life drawing, and naturism to casual adult selfies.
Finally, \lsexualfigurative{} label primarily contains anime, furry, or digitally rendered characters with exaggerated sexual features. These clusters often reflect niche online subcultures built around franchises like Pokémon, Marvel, and Final Fantasy, and include a strong commission-based creator economy (Patreon, YCH auctions). 
Compared to the other labels, sexual-figurative consists more of fictional or stylized bodies.

\noindent
\new{Similarly, Bluesky also issues action labels such as \lhide{}, \lwarn{} and \ltakedown{}, which enforce moderation decisions for all users. %
This also includes labels issued by country-specific moderation services. 
Most of the clusters in these action labels are centered around extreme political hostility and posts promoting conspiracies and extreme violence. 
For brevity and better readability, we place the detailed discussion on action labels and \lspam{} in Appendix~\ref{Sec: appendix actions}.
}

\subsection{Implications of the Observations} 

Across labels, our analyses point to potential correspondences between the observed landscape of harms and the systemic risks defined in DSA Article 34(1)~\cite{EU2022DSA}.

The clusters within \lintolerant{}, \lrude{}, and \lthreat{} -- esp., those involving hate speech -- may map to risks concerning \textit{illegal content} (some forms of hate speech) (Art. 34(1)(a)) and negative effects on \textit{fundamental rights} (Art. 34(1)(b)), particularly the rights to \textit{non-discrimination} and \textit{human dignity} and implicate risk to \textit{civic discourse} (Art. 34(1)(c)). 

Similarly, the \lgraphicmedia{} label, by identifying violent and gory imagery, hints at potential threats to \textit{public security} (Art. 34(1)(c)) and users' \textit{mental well-being} (Art. 34(1)(d)). 
Furthermore, the \lselfharm{} label suggests possible risks to \textit{physical and mental well-being} (Art. 34(1)(d)) through the potential normalization of self-injury. 

Sexual content labels such as \lsexual{}, \lporn{} could indicate risks related to the \textit{protection of minors} and \textit{gender-based violence} (Art. 34(1)(d)), particularly where content might involve exploitative or non-consensual imagery. 
While a granular post-level analysis to map specific content to DSA risk categories is outside the scope of the current study, our characterization of labeled content broadly hints at the existence of these potential systemic harms.

\section{Ethical considerations}\label{Sec: appendix ethics}

Our research is grounded in a commitment to ethical and responsible data analysis. 
The study exclusively uses publicly available data, analyzing social media posts that include text and images, and does not involve any non-public accounts related data, or direct user interactions. 
As such, this work does not constitute human-subjects research. 
Moreover, to protect the integrity and security of the data, all storage and processing were conducted on secure institutional infrastructure. 
Furthermore, our analytical approach was designed to minimize risk; we used in-house, open-source tools for our analyses. %
We did not interact with platform users, deploy automated agents, or manipulate platform behavior in any way. 
All results are presented using aggregated statistics. %
This careful methodology ensures that the positive impacts of %
the work substantially exceed any associated risks.

\section{Concluding Discussion}
\label{Sec: Discussion}

By leveraging the architectural transparency of the Bluesky decentralized platform, our work conducts the first systematic audit of a live moderation system (Bluesky Moderation Service), examining its mechanism, efficacy, and purpose. 
Our findings paint a detailed portrait of a human-AI collaborative system grappling with the classic trade-off between precision and recall--one that acts swiftly on clear-cut violations but relies on slower, nuanced human judgment for more complex harms. 
Finally, through unsupervised clustering, we reverse-engineered the de facto meaning of abstract policy labels, finding that their application in practice is often much broader than their written definitions suggest. This characterization also provides a detailed account of the types of harm detected by the Bluesky Moderation Service.
\neww{We contend that this study is itself evidence of the value of openness: the audit was only possible due to the architectural openness of the platform, which renders both the input data stream and moderation actions auditable, and our findings are in turn validated by the platform's own transparency reporting.}

\vspace{1mm} \noindent
\textbf{Limitations and future work: }
This study has several limitations that open avenues for future research. 
Our analysis was confined to Bluesky’s official moderation service; a crucial next step is to extend this audit to the broader federated ecosystem of third-party labelers to understand how they complement or contradict the default system. 
Furthermore, some of the methodologies--e.g., reliance on labeling delay for understanding human-AI collaboration--are proxy-dependent. \neww{Post-hoc data collection, particularly of media content, may also introduce some data loss: posts that are taken down between the moderation action and our collection window are not captured, which could skew estimates of the relative prevalence of harms.}
While our findings corroborate Bluesky's official transparency report, future work can delve into developing more robust methodologies. %
Finally, our work focused on the application of labels, but not their downstream effects on user behavior or content prevalence. 
\neww{Longitudinal studies are needed to measure these impacts and any temporal variations in moderation actions.}

\vspace{1mm} \noindent
\textbf{Takeaways for future moderation systems: }
Even with these limitations, our findings offer meaningful guidance to the development of better content moderation systems. 
First, our cluster-based analyses provide a data-driven roadmap to improve the current labeling paradigm through increased label granularity. \neww{Second, although identifying unsafe content is a challenge all digital platforms grapple with, they largely do so independently. Moreover, even with regulatory transparency requirements such as those under the DSA, moderation practices remain opaque: the most common reason reported in the DSA Transparency Database is the uninformative ``Other violation of provider’s terms and services.''
}
To this end, the research community could draw inspiration from the collaborative efforts in spam analytics to develop joint initiatives for moderation analytics by adopting open standards like those on Bluesky. 
We believe our study is an initial step in this direction.

\section*{Acknowledgments}
Ingmar Weber is supported by funding from the Alexander von Humboldt Foundation and its founder, the Federal Ministry of Education and Research (Bundesministerium für Bildung und Forschung). Abhijnan Chakraborty acknowledges the funding support from the Max Planck Society. 

\bibliography{main}

\appendix
\section{Bluesky Moderation Service: Statistics}\label{Sec: appendix stats other labels}

\begin{table*}[t]
\centering
\caption{\new{Label statistics: available records, embed-type distribution,
root post vs.\ reply breakdown, and top-3 languages per label.
Here, thumb=external link thumbnail,
record=quoted post, record+media=quoted post with media.
Language tags normalised to BCP-47 base; posts with no tag counted as \textit{no-lang}.}}
\small
\label{tab:label_stats_combined}
\resizebox{\textwidth}{!}{%
\begin{tabular}{lrrrrrrrrrl}
\toprule
\multicolumn{1}{l}{\textbf{Label}} &
\multicolumn{1}{c}{\textbf{Records}} &
\multicolumn{6}{c}{\textbf{Embed type}} &
\multicolumn{2}{c}{\textbf{Root / Reply}} &
\multicolumn{1}{l}{\textbf{Top langs}} \\
\cmidrule(lr){3-8} \cmidrule(lr){9-10}
 & & \textbf{thumb} & \textbf{image(s)} & \textbf{record} & \textbf{record+media} & \textbf{video} & \textbf{text-only} & \textbf{root} & \textbf{reply} & \\
\midrule
porn & 7,360,828 & 415,280 & 4,907,245 & 171 & 100,400 & 1,937,717 & 15 & 6,704,119 (91\%) & 656,709 (9\%) & en (69\%), no-lang (11\%), es (4\%) \\
sexual & 2,243,593 & 170,884 & 1,632,310 & 33 & 41,808 & 398,541 & 17 & 2,061,542 (92\%) & 182,051 (8\%) & en (65\%), no-lang (14\%), ja (4\%) \\
nudity & 271,246 & 17,515 & 226,590 & 21 & 5,445 & 21,669 & 6 & 248,942 (92\%) & 22,304 (8\%) & en (68\%), no-lang (12\%), es (4\%) \\
rude & 219,712 & 4,007 & 8,860 & 5,700 & 627 & 284 & 200,234 & 8,502 (4\%) & 211,210 (96\%) & en (97\%), fr (1\%), de (1\%) \\
!takedown & 214,079 & 1,409 & 8,931 & 1,946 & 420 & 1,691 & 199,682 & 195,252 (91\%) & 18,827 (9\%) & no-lang (84\%), en (14\%), pt (0\%) \\
sexual-figurative & 134,942 & 961 & 121,184 & 23 & 4,861 & 7,904 & 9 & 126,611 (94\%) & 8,331 (6\%) & en (81\%), ja (7\%), no-lang (4\%) \\
graphic-media & 93,788 & 28,229 & 39,179 & 8 & 4,117 & 22,252 & 3 & 72,853 (78\%) & 20,935 (22\%) & en (64\%), no-lang (11\%), es (7\%) \\
spam & 78,077 & 24,143 & 4,775 & 3,543 & 113 & 212 & 45,291 & 711 (1\%) & 77,366 (99\%) & en (71\%), ja (5\%), pt (4\%) \\
intolerant & 38,021 & 997 & 3,723 & 1,988 & 180 & 185 & 30,948 & 10,550 (28\%) & 27,471 (72\%) & en (90\%), ja (2\%), es (1\%) \\
threat & 11,640 & 296 & 677 & 682 & 58 & 44 & 9,883 & 2,408 (21\%) & 9,232 (79\%) & en (93\%), fr (1\%), pt (1\%) \\
self-harm & 10,187 & 388 & 8,108 & 7 & 373 & 1,088 & 223 & 8,944 (88\%) & 1,243 (12\%) & en (49\%), pt (39\%), ja (3\%) \\
!hide & 3,520 & 298 & 483 & 236 & 32 & 11 & 2,460 & 1,520 (43\%) & 2,000 (57\%) & en (92\%), no-lang (5\%), es (2\%) \\
!warn & 2,191 & 68 & 526 & 184 & 22 & 6 & 1,385 & 1,090 (50\%) & 1,101 (50\%) & en (96\%), de (1\%), no-lang (0\%) \\
\midrule
\textbf{Total} & 10,681,824 & 664,475 & 6,962,591 & 14,542 & 158,456 & 2,391,604 & 490,156 & 9,443,044 (88\%) & 1,238,780 (12\%) & \\
\bottomrule
\end{tabular}}
\vspace{-4mm}
\end{table*}

\begin{table}[b]
\centering
\caption{Top labels applied on accounts and profiles by the BMS in 2025.}
\small
\label{tab:other_label_counts}
\begin{tabular}{ll||ll}
\toprule
\textbf{Account Labels} & \textbf{Count} & \textbf{Profile Labels} & \textbf{Count} \\
\midrule
!takedown      & 2{,}040{,}503 & porn              & 63{,}969 \\
needs-review   &   213{,}927   & nudity            & 13{,}859 \\
spam           &   113{,}576   & sexual            &  3{,}827 \\
!hide          &    43{,}417   & sexual-figurative &  1{,}398 \\
impersonation  &     3{,}185   & graphic-media     &    859   \\
rude           &     1{,}179   & !takedown         &    347   \\
intolerant     &       828     & self-harm         &    295   \\
!warn          &       656     & spam              &     28   \\
\bottomrule
\end{tabular}
\vspace{-4mm}
\end{table}

\Cref{tab:label_stats_combined} gives statistics about labeled records collected and their distribution across label values, post embed types and languages. 
\Cref{tab:other_label_counts} shows the top labels applied to accounts and profiles by BMS.  %
For account labels, \ltakedown{} is the most frequently applied label, whereas \lporn{} is the most frequent label on profiles. 
When a label is applied to an account, it will show warnings on all of its content. 
On the other hand, a label on a profile is usually used to blur its avatar without showing any warnings on its posts.

\section{Additional Experimental Details for RQ2: Firehose Collection and Indexing}\label{Sec: appendix firehose}

\noindent\new{\textbf{Firehose Data Collection: }Using the \texttt{com.atproto.sync.subscribeRepos} endpoint, we collected firehose events between March and December 2025, yielding 1.14B post records (events of type \texttt{app.bsky.feed.post}), of which 11.9M ($\sim$1\%) posts were labeled by the BMS. Each post record contains fields such as the post text, creation timestamp, and optional embedded media content identifiers (CIDs). From this collection, we derive two datasets used in our experiments. For the recall evaluation (Section~\ref{Sec: Accuracy}), we randomly sample 1,000 posts. For experiment in (Section~\ref{sec:blindspots}), we curate a subset of 40M posts drawn randomly from the March--June 2025 window (383.2M posts total), restricted to text and single-image posts.}

\noindent\new{\textbf{Embedding, Indexing and Nearest-Neighbour Retrieval.} For each post in the 40M subset, we compute multimodal embeddings using \texttt{Qwen3-VL-Embedding-2B}~\cite{li2026qwen3}\footnote{\url{https://huggingface.co/Qwen/Qwen3-VL-Embedding-2B}}. Each post is encoded as a multimodal instruction, combining available text and a single image into a 2048 dimensional vector. All embeddings are normalised and inserted sequentially into a \texttt{faiss.IndexFlatIP} index of same dimension. Embeddings for the 336 labeled posts (described in Section~\ref{sec:blindspots}) are searched against the firehose index using k=1000 candidates per query. For each labeled post, we walk the ranked neighbor list and select the highest-ranked neighbor that (a) carries no BMS label and (b) contains exactly one image, yielding one semantically similar but unmoderated post per labeled query i.e set of 336 unlabeled posts used in Section~\ref{sec:blindspots}.}

\section{Automod and Hive AI Moderation}\label{Sec: appendix hive}
\begin{figure}
    \centering
    \includegraphics[width=0.8\columnwidth, height=3.5cm]{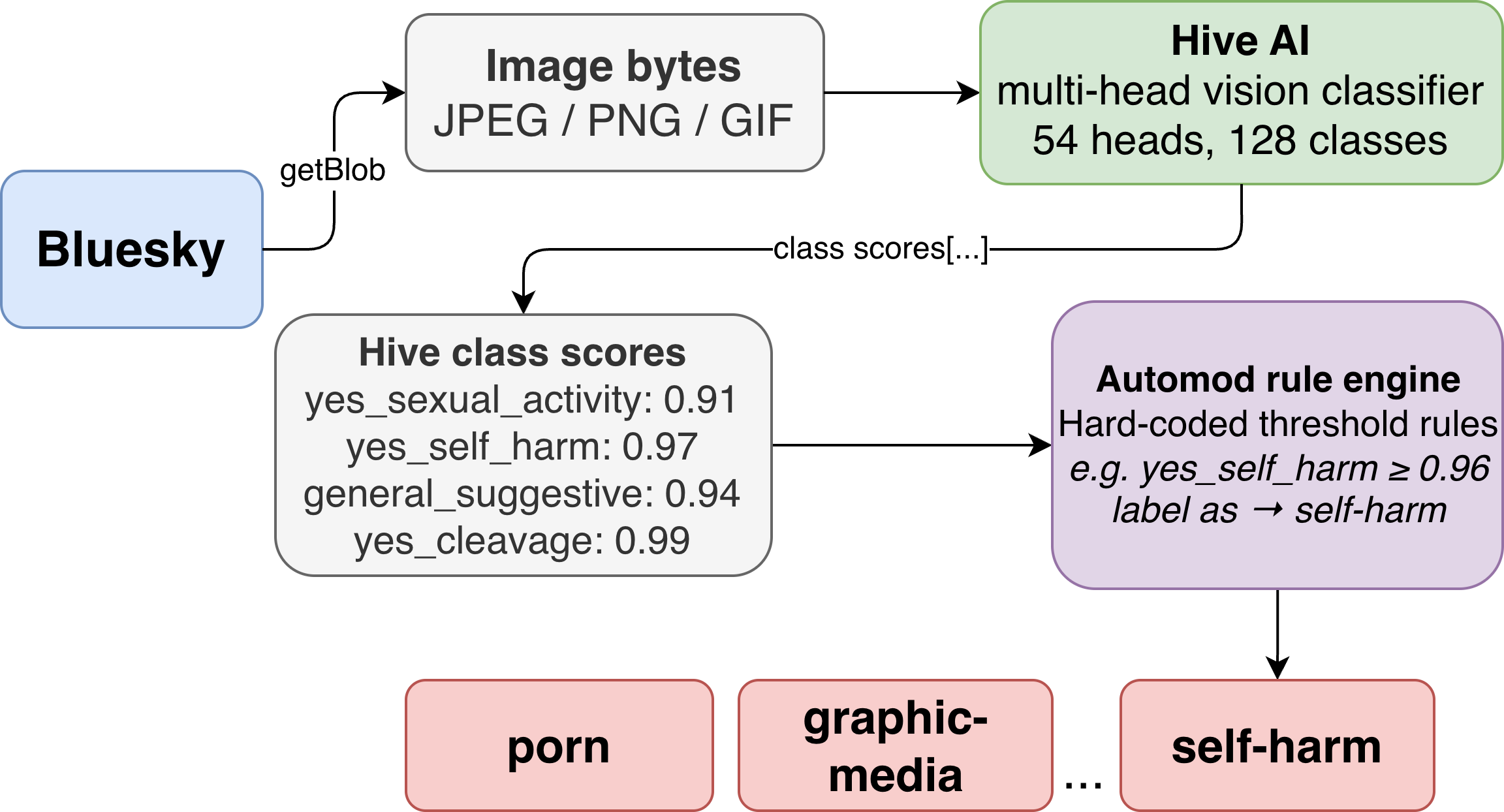}
    \caption{\new{Pipeline of Automod. A post image is submitted to the Hive AI classifier, which returns scores across 128 classes and 54 heads. Automod applies hard-coded threshold rules over 16 of these classes to produce BMS labels.}}
    \label{fig:hivepipe}
\end{figure}

\noindent\new{\textbf{Overview.} BMS's automated content moderation is powered by two components operating in sequence: \textit{Hive AI}, a commercial multi-head vision classifier, and \textit{Automod}, BMS's open-source rule engine that converts Hive's raw scores into actionable labels. Figure~\ref{fig:hivepipe} illustrates this pipeline.}

\noindent\new{\textbf{Hive AI.} When a post containing an image (individual frames in case of video) is submitted, BMS submits it to the Hive  API. Hive AI's visual moderation API returns a flat list of 128 class-score pairs organised into 54 model heads spanning five content domains: sexual content (26 heads, 59 classes), violence and gore (10 heads, 29 classes), drugs and vices (6 heads, 15 classes), hate imagery (5 heads, 10 classes), and miscellaneous image attributes (7 heads, 15 classes), as documented in the Hive Visual Moderation API~\cite{hiveai}. Within each head, classes are mutually exclusive, and their scores sum to 1. For example, the \textit{Sexual Activity} head returns \texttt{yes\_sexual\_activity} and \texttt{no\_sexual\_activity}; the \textit{Blood} head returns \texttt{very\_bloody}, \texttt{a\_little\_bloody}, \texttt{other\_blood}, and \texttt{no\_blood}. Table~\ref{tab:automod_heads} lists the heads and classes relevant to BMS's five automated labels. Table~\ref{tab:h2_heads} lists a few heads outside Automod's rules.}

\vspace{1mm}
\begin{table}[ht]
\centering
\small
\setlength{\tabcolsep}{4pt}
\renewcommand{\arraystretch}{0.95}
\caption{\new{Hive classes read by Automod rules.}} 
\label{tab:automod_heads}
\resizebox{\columnwidth}{!}{
\begin{tabular}{ll}
\toprule
\textbf{Hive class} & \textbf{Description} \\
\midrule
\texttt{yes\_sexual\_activity}          & Sex acts or genital stimulation \\
\texttt{yes\_realistic\_nsfw}           & Live or photo-realistic nudity/sex acts \\
\texttt{general\_nsfw}                  & Genitalia, sexual activity, nudity, buttocks, sex toys \\
\texttt{yes\_sexual\_intent}            & Occluded, blurred, or hidden sexual activity \\
\texttt{yes\_sex\_toy}                  & Dildos or certain lingerie \\
\texttt{yes\_male\_nudity}              & Male genitalia \\
\texttt{yes\_female\_nudity}            & Breasts or female genitalia \\
\texttt{yes\_undressed}                 & Nude even if genitals occluded by pose or overlay \\
\texttt{yes\_male\_underwear}           & Male underwear \\
\texttt{yes\_female\_underwear}         & Female underwear \\
\texttt{yes\_self\_harm}                & Self-cutting, burning, or suicide methods  \\
\texttt{very\_bloody}                   & Gore or visible bleeding \\
\texttt{human\_corpse}                  & Human dead body present \\
\texttt{hanging}                        & Human hanging by noose \\
\bottomrule
\end{tabular}}
\end{table}

\vspace{1mm}
\begin{table}[ht]
\centering
\small
\setlength{\tabcolsep}{4pt}
\renewcommand{\arraystretch}{0.95}
\caption{\new{Hive classes outside Automod's rule set --- H2 blindspot heads.}}
\resizebox{\columnwidth}{!}{
\begin{tabular}{ll}
\toprule
\textbf{Hive class} & \textbf{Description} \\
\midrule
\texttt{general\_suggestive}    & \parbox[t]{5cm}{Shirtless men, underwear/swimwear,\\suggestive poses without genitalia} \\[4pt]
\texttt{yes\_cleavage}          & Identifiable female cleavage \\
\texttt{yes\_male\_shirtless}   & Shirtless below mid-chest \\
\texttt{yes\_bulge}             & Penis visible underneath clothing \\
\texttt{yes\_female\_swimwear}  & Bikinis, one-pieces \\
\texttt{yes\_bodysuit}          & Bodysuits not covering the thigh \\
\bottomrule
\end{tabular}}
\label{tab:h2_heads}
\end{table}
\vspace{1mm}

\noindent\new{\textbf{Automod rule engine.} Automod applies hard-coded threshold rules over Hive's output scores across the different classes to produce BMS labels. Each rule checks a specific Hive class against a fixed threshold: for example, \texttt{yes\_self\_harm} $\geq$ 0.96 triggers \lselfharm{}, and \texttt{yes\_sexual\_activity} $\geq$ 0.90 triggers \lporn{}. Sexual content labels follow a priority cascade (\lporn{} before \lsexual{} before \lnudity{}). In total, Automod acts on 16 classes across 10 heads out of 128 classes and 54 heads total; the remaining 112 classes across 44 heads are not used. This design gives rise to both recall failure modes identified in Section~\ref{sec:blindspots}.}

\section{Human Oversight in Redressal and Country Specific Moderation}\label{Sec: appendix delays}

\begin{figure}[t]
    \centering
    \begin{subfigure}[t]{0.48\columnwidth}
    \centering
    \includegraphics[width=\linewidth, height=3cm]{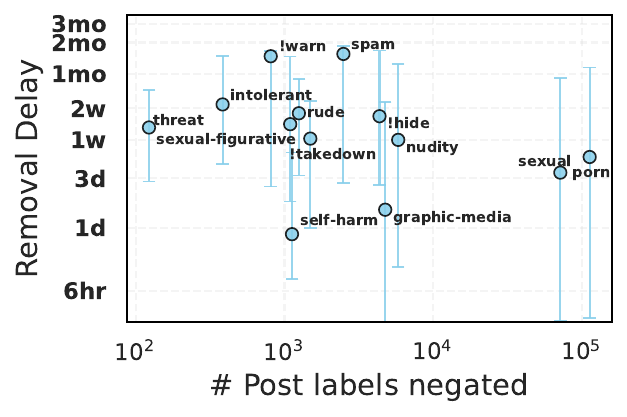}
    \caption{Redressal delays}
    \label{fig:delay_neg}
  \end{subfigure}
  \begin{subfigure}[t]{0.48\columnwidth}
    \centering
    \includegraphics[width=\linewidth, height=3cm]{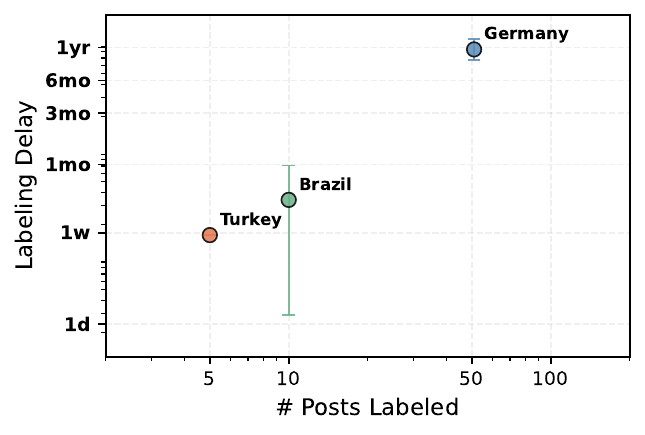}
    \caption{Delay for country specific labeler}
    \label{fig:delay_country}
  \end{subfigure}
  \vspace{-2mm}
  \caption{(a)~Redressal Latency: The median delay between post creation and the negation (removal) of a label. (b)~Labeling delays for country-specific moderation (Germany, Brazil, Turkey).}
  \label{fig:delays_bc}
  \vspace{-4mm}
\end{figure}

\subsection{Country-specific Moderation Services: } As shown in Figure~\ref{fig:delay_country}, country-specific moderation services operate on a fundamentally different timescale and scale compared to the BMS. The labeling delays (\lhide{} label) are often exceptionally high, ranging from days to weeks (Turkey, Brazil) to nearly a year (Germany) for a few posts. This contrast from the default moderation service suggests that country-specific services do not function as a proactive safety system but rather as a reactive compliance mechanism.

\subsection{Negation labels \& Redressal mechanism: }
BMS also often negates or reverses its labels. 
A negation label functions as an explicit override, i.e., it cancels or reverses a previously assigned label.
\new{We conduct a similar analysis on negation labels
across 13 label categories, covering 206,724 negated post labels
in total. The median delay between post creation and label
negation varies substantially across label types, as shown in
Figure~\ref{fig:delay_neg}. Action labels (\lwarn{}, \lhide{}) and labels such as \lspam{}, \lintolerant{} and \lrude{} show the longest removal delays, with
median delays of 44–47 days for \lspam{} and \lwarn{}, and 12–15
days for \lrude{} and \lintolerant{}. In contrast, content labels such
as \lselfharm{} (21h), \lgraphicmedia{} (1.5d), and \lsexual{} (3.4d)
are negated considerably faster.}
Most of the negated labels are potentially automated, e.g., \lporn{}, \lsexual{}, \lnudity{}, etc., indicating these labels might have been automatically (but erroneously) applied to the posts and the authors of the posts may have reported such mislabeling through Bluesky's redressal mechanism.   

\section{Clustering labeled posts}\label{Sec: appendix clusters}
\begin{table}[t]
\centering

\caption{Optimal hyperparameters for clustering each moderation label and the corresponding DBCV scores.}
\label{tab:hyperparam}

\resizebox{\linewidth}{!}{%
\begin{tabular}{lccccrrr}
\toprule
Label & $n\_neighbors$ & $min\_cluster\_size$ & $min\_samples$ & DBCV & Total Posts & Noise Posts & Noise \%\\
\midrule
sexual & 37 & 4179 & 540 & 0.626 & 300,000 & 88,783 & 29.6\% \\
porn & 32 & 4815 & 85 & 0.407 & 300,000 & 159,110 & 53.0\% \\
nudity & 38 & 1106 & 190 & 0.371 & 268,376 & 122,229 & 45.5\% \\
sexual-figurative & 35 & 1065 & 48 & 0.390 & 134,855 & 71,345 & 52.9\% \\
graphic-media & 43 & 1164 & 160 & 0.751 & 93,313 & 17,228 & 18.5\% \\
intolerant & 21 & 299 & 27 & 0.696 & 36,023 & 8,233 & 22.9\% \\
rude & 27 & 2706 & 46 & 0.387 & 214,001 & 105,267 & 49.2\% \\
self-harm & 16 & 139 & 23 & 0.566 & 10,169 & 3,557 & 35.0\% \\
threat & 16 & 222 & 24 & 0.693 & 10,958 & 2,246 & 20.5\% \\
spam & 38 & 1095 & 25 & 0.541 & 74,521 & 18,922 & 25.4\% \\
!warn & 29 & 107 & 1 & 0.910 & 2,006 & 0 & 0.0\% \\
!hide & 25 & 50 & 4 & 0.764 & 3,278 & 54 & 1.6\% \\
!takedown & 25 & 674 & 3 & 0.885 & 31,129 & 860 & 2.8\% \\
\bottomrule
\end{tabular}%
}
\vspace{-2mm}
\end{table}

\begin{table}[t]
\centering
\scriptsize
\caption{\new{Prompt used for the chunk-level summarisation with \texttt{Qwen3-VL-32B-Instruct} (map stage).}}
\label{tab:prompt_chunk}
\begin{tabular}{p{0.9\linewidth}}
\toprule
\textbf{User Prompt:} \\
\midrule
Summarize these social media posts in 2--3 sentences.
Focus on shared themes, sentiment, and any important visual context. \\[2mm]
{[POST\_START 1]} \\
\texttt{<text of post 1>} \\
\texttt{<image(s) of post 1, if any>} \\
\texttt{<video frames of post 1, if any>} \\
{[POST\_END]} \\[1mm]
{[POST\_START 2]} \\
\texttt{<text of post 2>} \\
\texttt{<image(s) of post 2, if any>} \\
\texttt{<video frames of post 2, if any>} \\
{[POST\_END]} \\
\(\vdots\) \\
\bottomrule
\end{tabular}
\end{table}

\begin{table}[t]
\centering
\scriptsize
\caption{\new{Prompt used for the reduce stage of map-reduce summarisation (combining chunk-level summaries into a final cluster description).}}
\label{tab:prompt_reduce}
\begin{tabular}{p{0.9\linewidth}}
\toprule
\textbf{User Prompt:} \\
\midrule
Combine these summaries into ONE final output of 1--2 sentences. \\
Capture overall themes, sentiment, and major visual trends. \\[2mm]
\texttt{<chunk summary 1>} \\
\texttt{<chunk summary 2>} \\
\(\vdots\) \\
\bottomrule
\end{tabular}
\end{table}

\begin{table}[t]
\centering
\scriptsize
\caption{\new{Prompt used for cluster naming with \texttt{Qwen3-VL-32B-Instruct}.
         All cluster summaries for a label are presented jointly so names reflect
         within-label distinctions.}}
\label{tab:prompt_naming}
\begin{tabular}{p{0.9\linewidth}}
\toprule
\textbf{System Prompt:} \\
\midrule
You are an expert content-taxonomy analyst.
You will receive a list of cluster summaries --- all clusters belong to the same
parent label. Your task is to assign each cluster a short, contrastive name
(2--5 words) that: \\[1mm]
\quad 1. Clearly signals what makes that cluster different from the others. \\
\quad 2. Stays specific and avoids generic phrases like ``related content'' or ``posts about''. \\
\quad 3. Is understandable by a non-specialist moderator. \\[2mm]
Respond only with a JSON object mapping cluster\_id (integer key as string) to
cluster\_name (string).
No explanation, no markdown fences, no extra keys. \\
Example format: \texttt{\{"0": "Racial slurs in sport", "1": "Anti-immigrant rhetoric", ...\}} \\
\midrule
\textbf{User Prompt:} \\
\midrule
Parent label: \texttt{<label>} \\[1mm]
Below are the cluster summaries (one per cluster): \\[1mm]
--- CLUSTER 0 \quad (n=\texttt{<n>}) --- \\
\texttt{<cluster 0 summary>} \\[1mm]
--- CLUSTER 1 \quad (n=\texttt{<n>}) --- \\
\texttt{<cluster 1 summary>} \\
\(\vdots\) \\[2mm]
Now generate a short, contrastive name for every cluster listed above.
Return only the JSON object described in the system prompt. \\
\bottomrule
\end{tabular}
\end{table}

\begin{figure*}
  \centering

  \begin{subfigure}[t]{0.33\textwidth}
    \centering
    \includegraphics[width=\linewidth, height = 3.5cm]{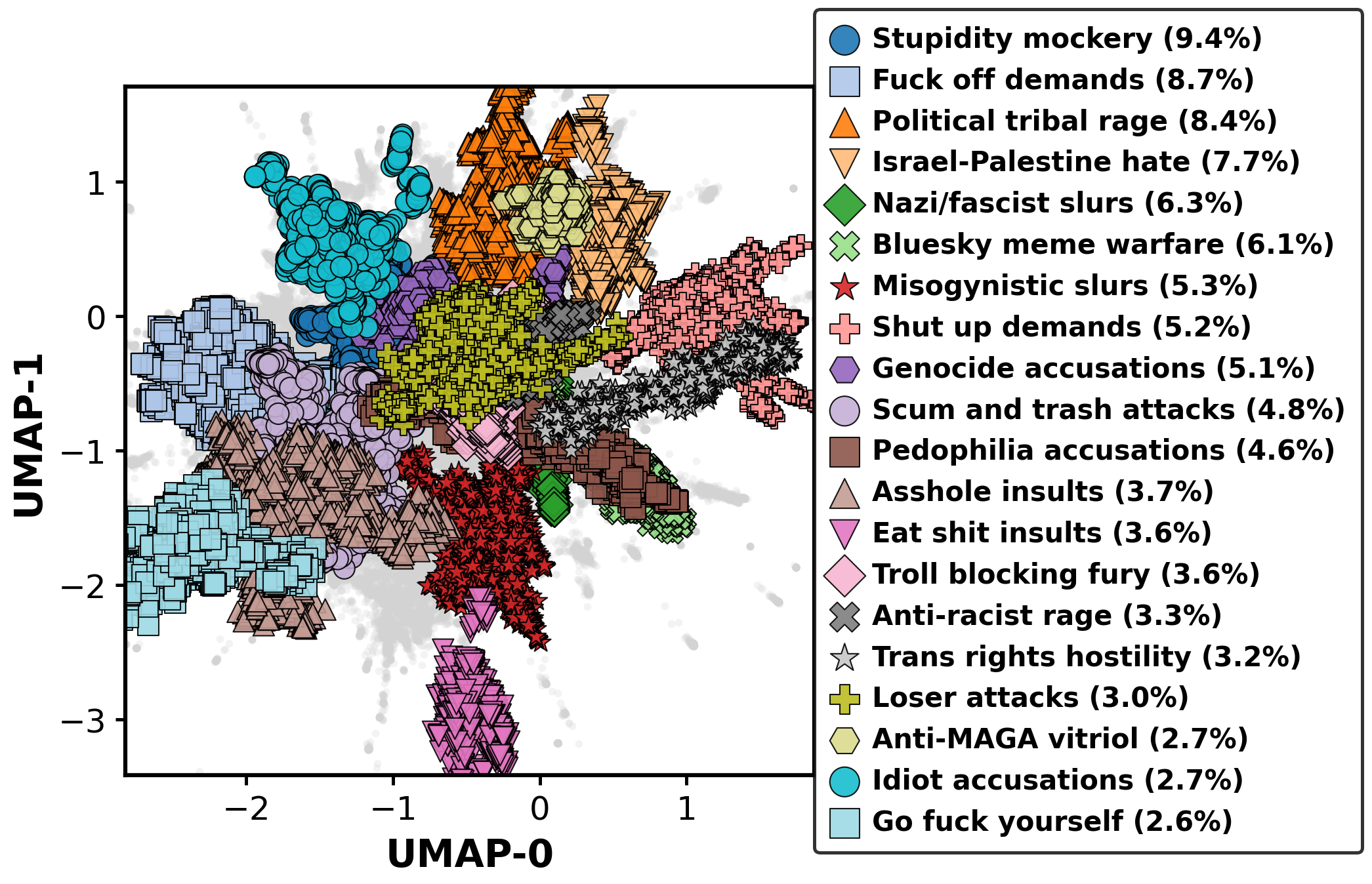}
    \caption{Rude}
    \label{fig:rudeA}
  \end{subfigure}
  \hfill
  \begin{subfigure}[t]{0.33\textwidth}
    \centering
    \includegraphics[width=\linewidth, height = 3.5cm]{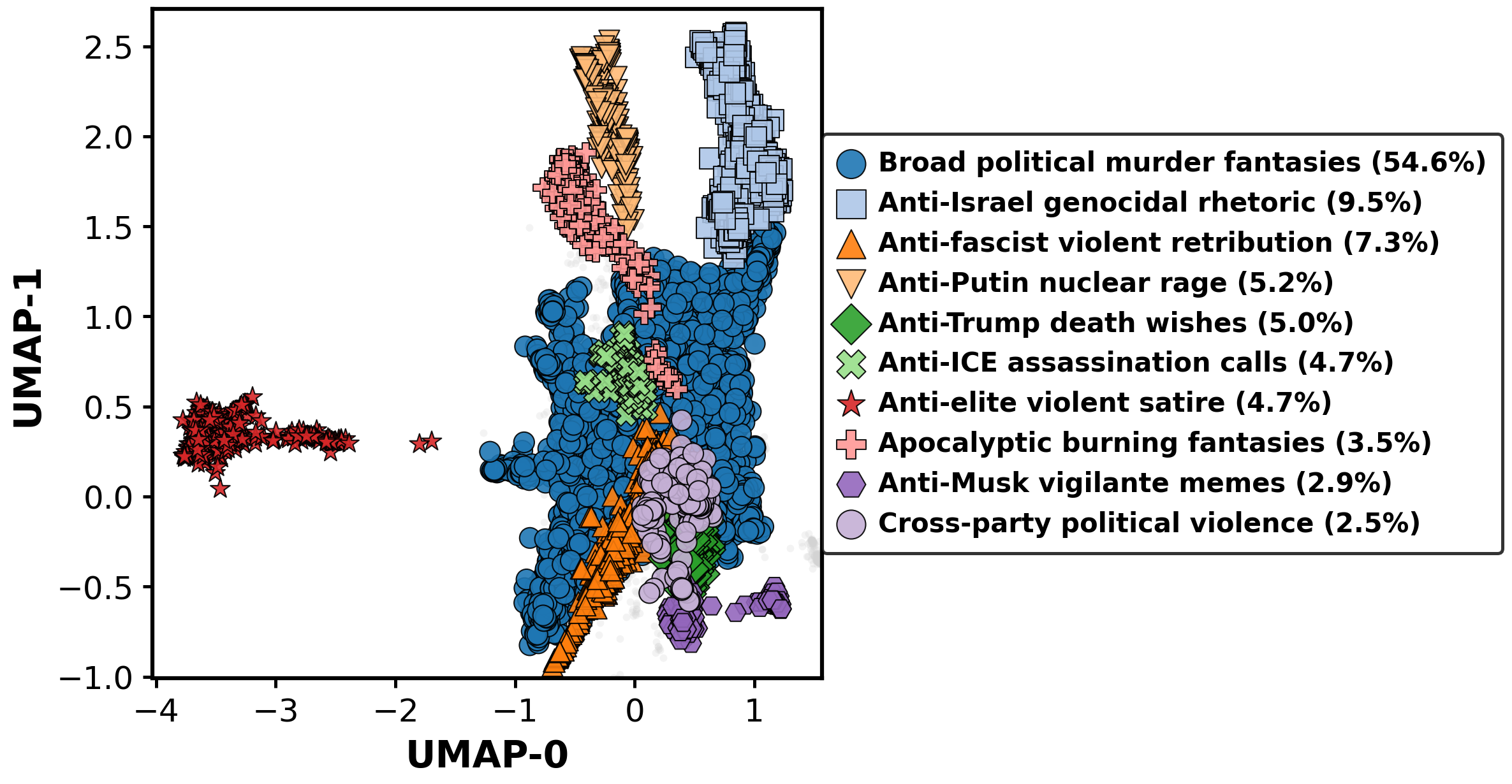}
    \caption{Threat}
    \label{fig:threatb}
  \end{subfigure}
  \hfill
  \begin{subfigure}[t]{0.33\textwidth}
    \centering
    \includegraphics[width=\linewidth, height = 3.5cm]{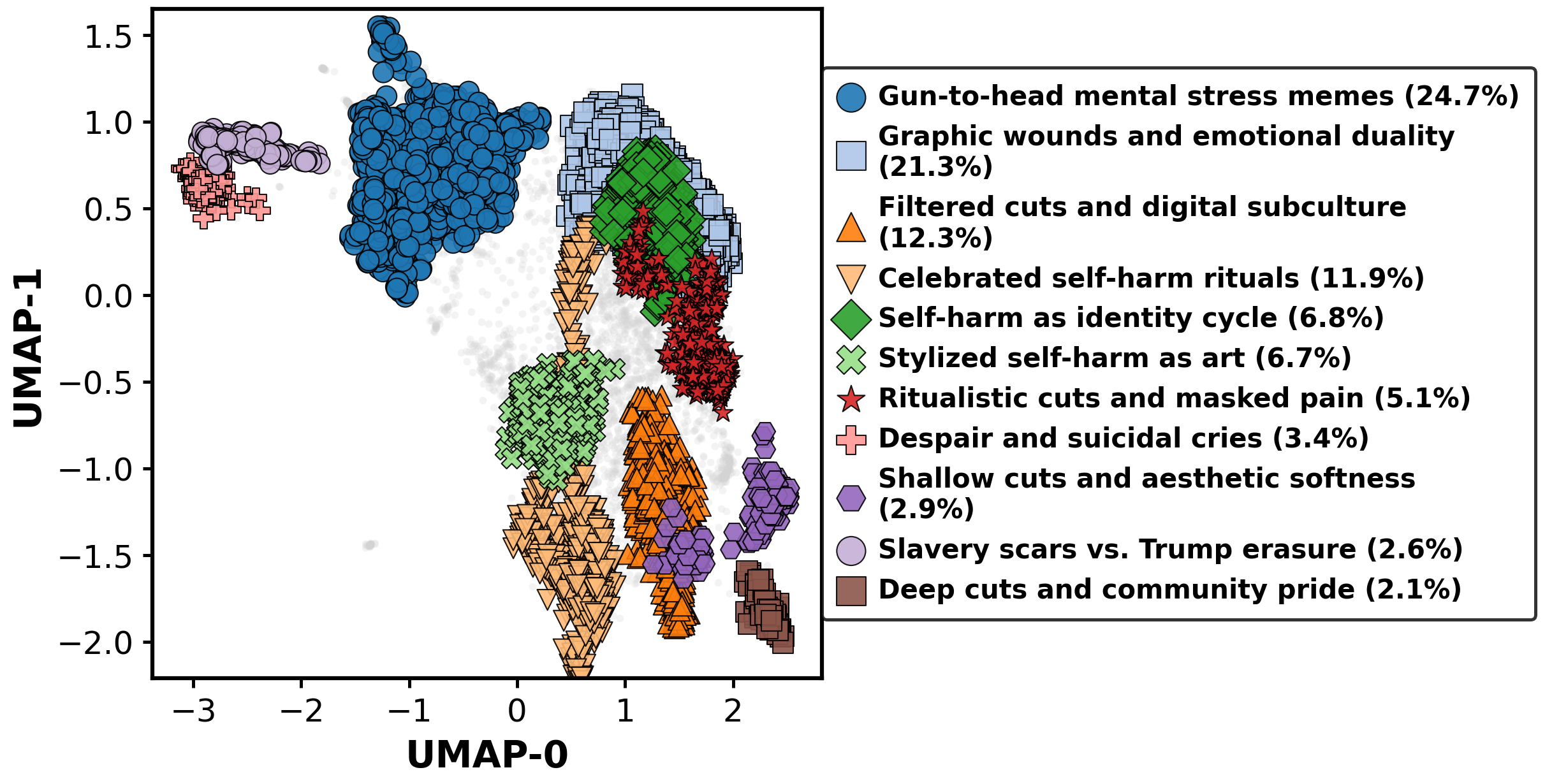}
    \caption{Self-harm}
    \label{fig:selfharmc}
  \end{subfigure}

  \vspace{0.4cm} %

  \begin{subfigure}[t]{0.3\textwidth}
    \centering
    \includegraphics[width=\linewidth, height = 3.5cm]{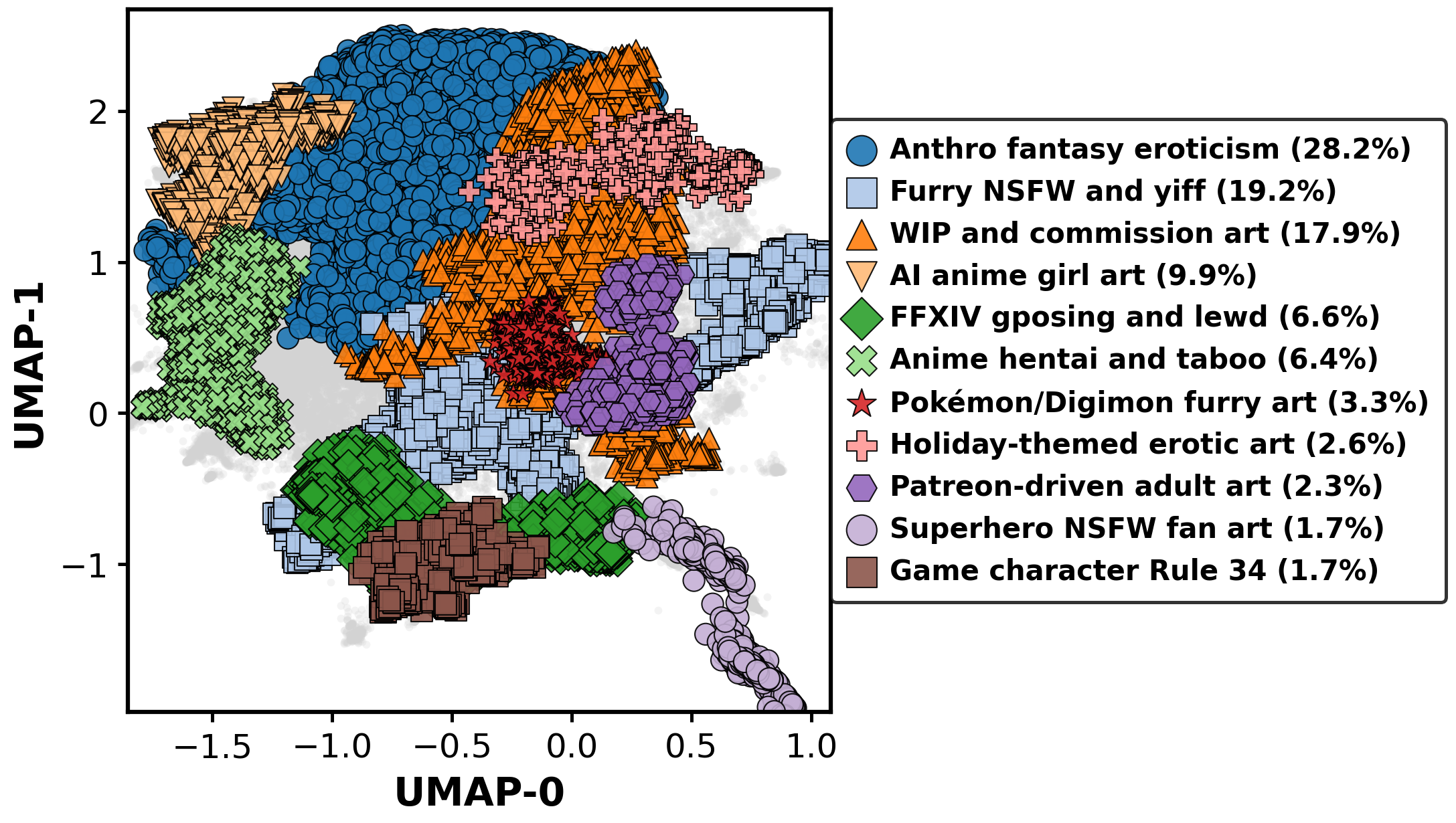}
    \caption{Sexual-figurative}
    \label{fig:sexualfigd}
  \end{subfigure}
  \hfill
  \begin{subfigure}[t]{0.3\textwidth}
    \centering
    \includegraphics[width=\linewidth, height = 3.5cm]{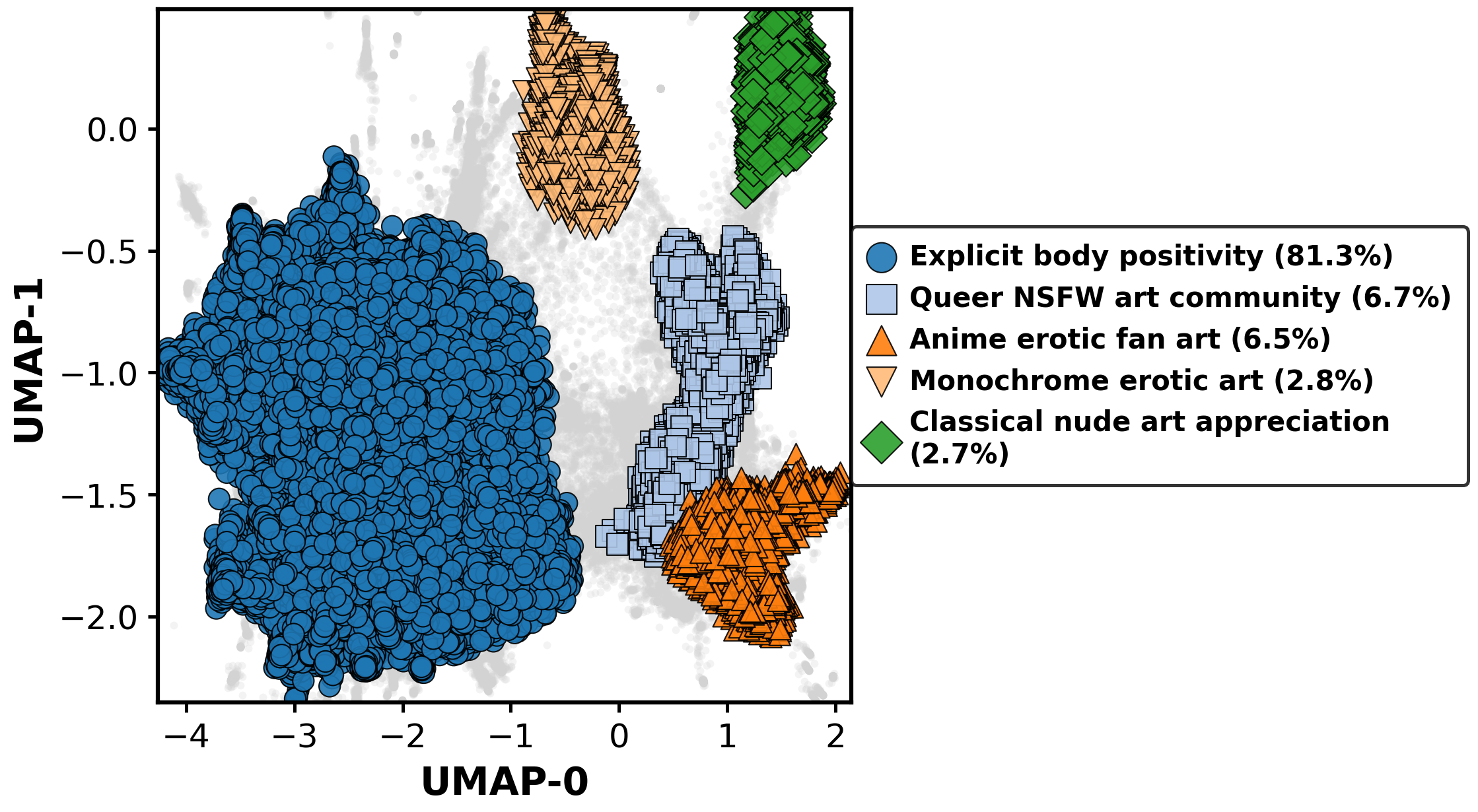}
    \caption{Sexual}
    \label{fig:sexuale}
  \end{subfigure}
  \hfill
  \begin{subfigure}[t]{0.3\textwidth}
    \centering
    \includegraphics[width=\linewidth, height = 3.5cm]{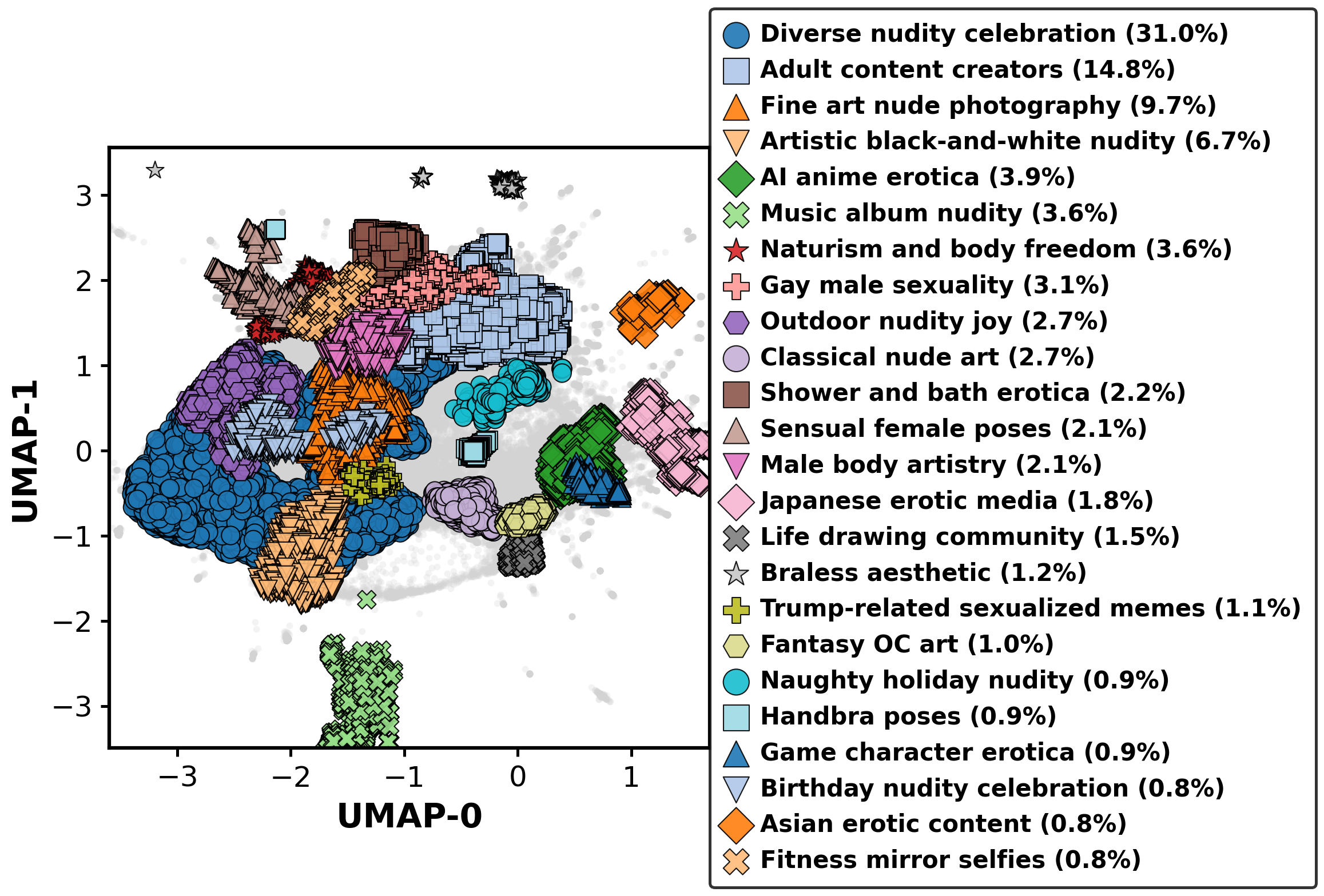}
    \caption{Nudity}
    \label{fig:nudityf}
  \end{subfigure}

  \begin{subfigure}[t]{0.3\textwidth}
    \centering
    \includegraphics[width=\linewidth, height=3.5cm]{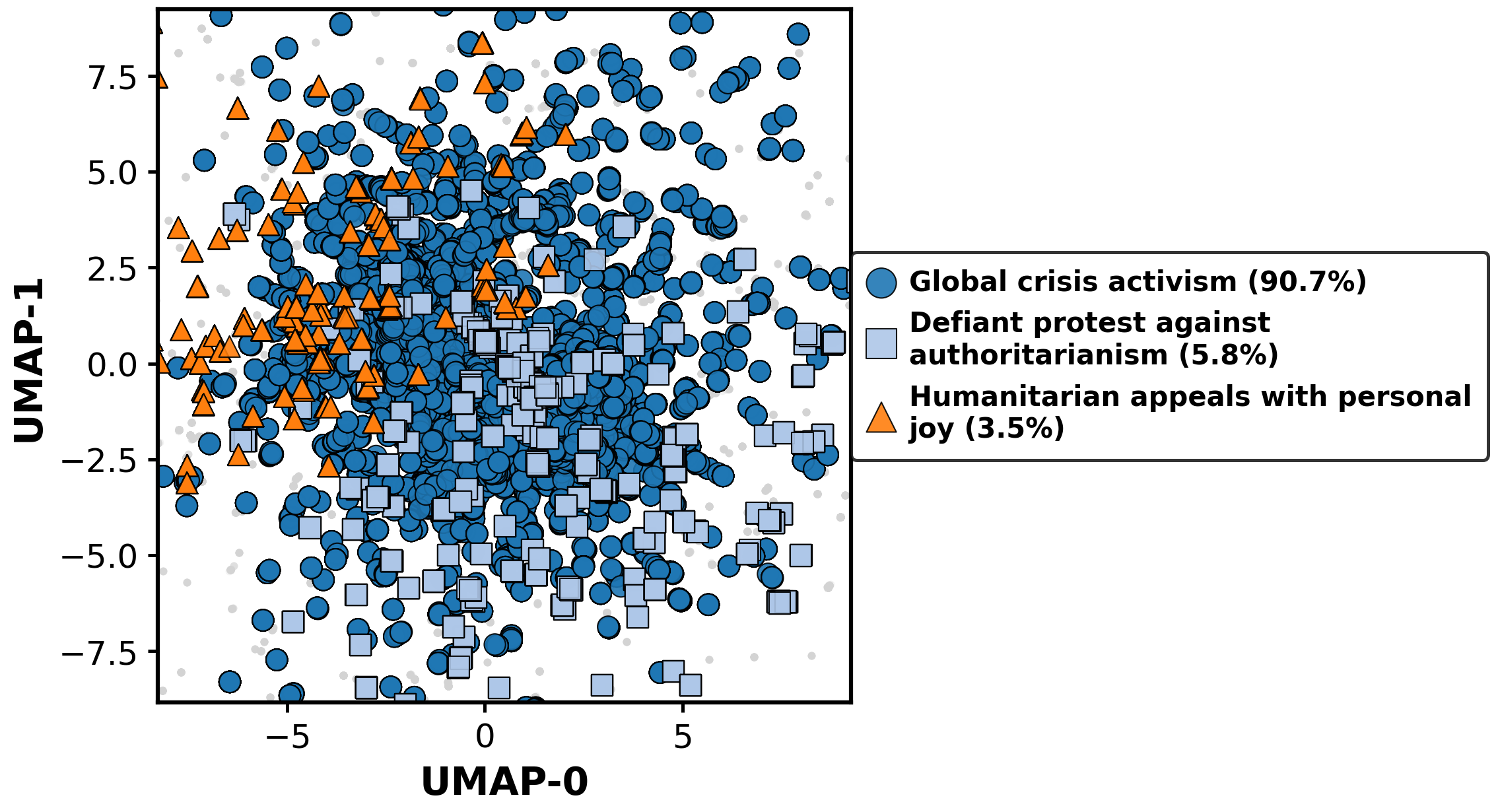}
    \caption{Spam}
    \label{fig:spamg}
  \end{subfigure}
  \hfill
  \begin{subfigure}[t]{0.3\textwidth}
    \centering
    \includegraphics[width=\linewidth, height=3.5cm]{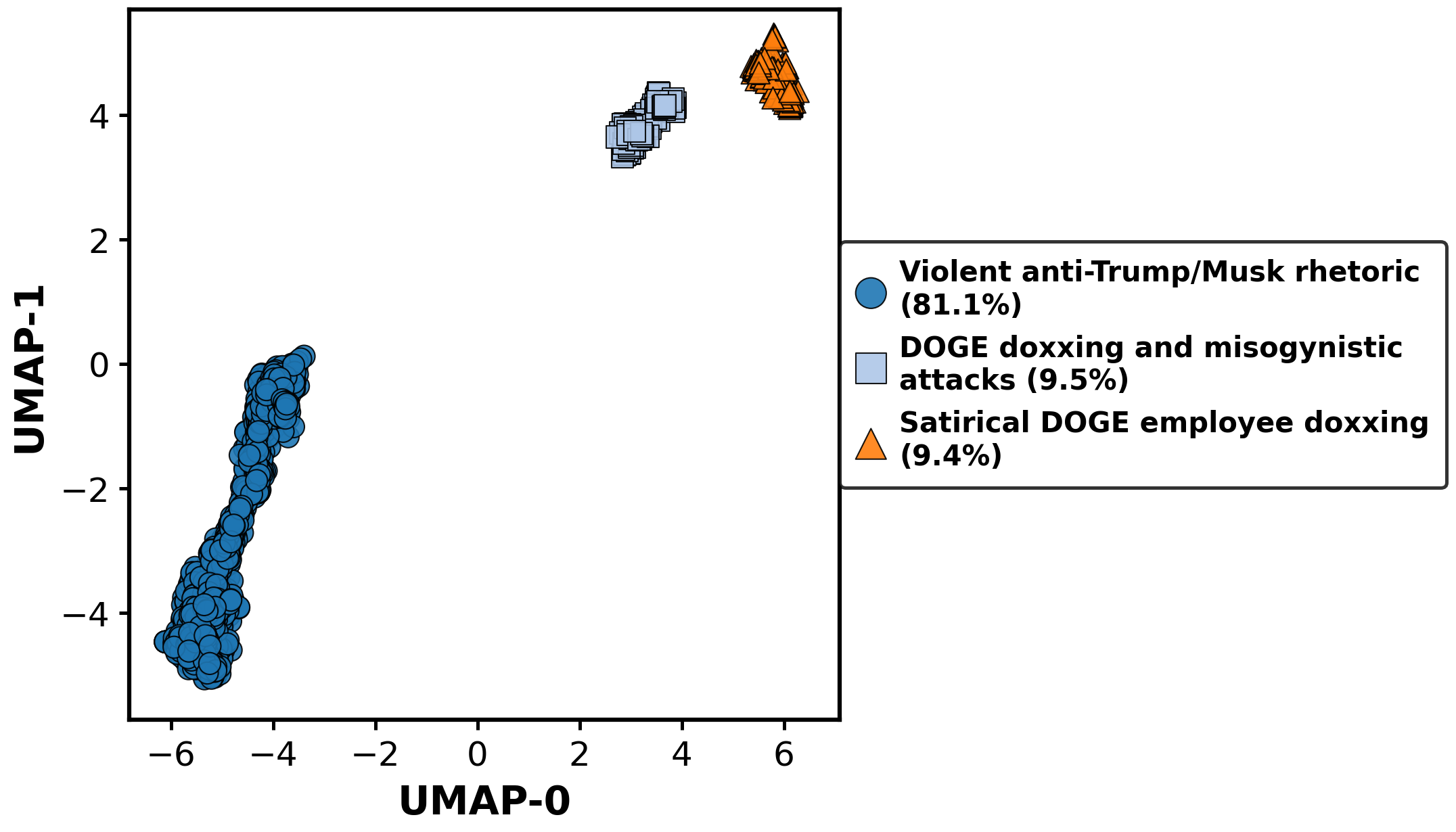}
    \caption{!warn}
    \label{fig:warnh}
  \end{subfigure}
  \hfill
  \begin{subfigure}[t]{0.3\textwidth}
    \centering
    \includegraphics[width=\linewidth, height=3.5cm]{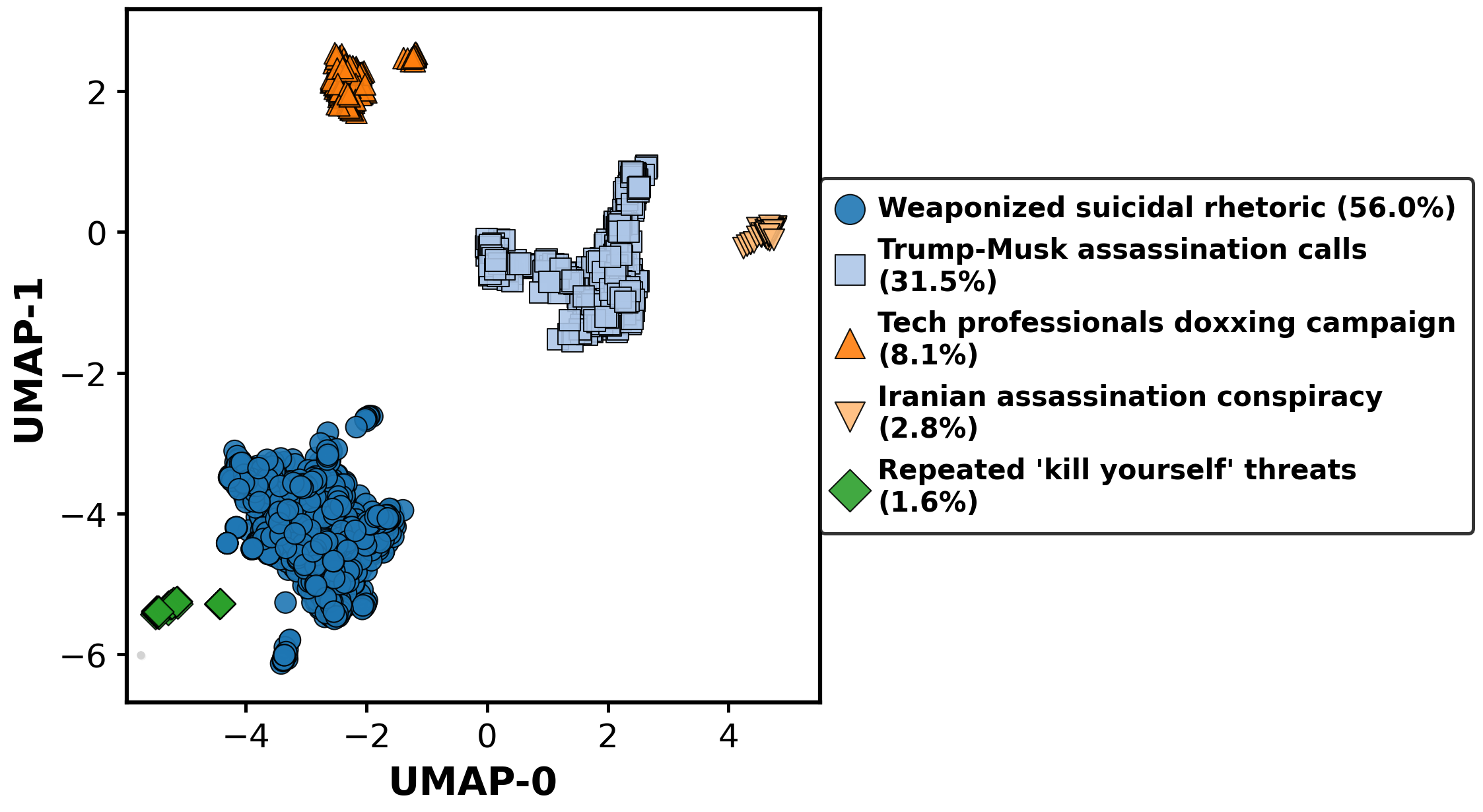}
    \caption{!hide}
    \label{fig:hidei}
  \end{subfigure}

  \vspace{0.4cm}

\begin{center}
  \begin{subfigure}[t]{0.32\textwidth}
    \centering
    \includegraphics[width=\linewidth, height=3.5cm]{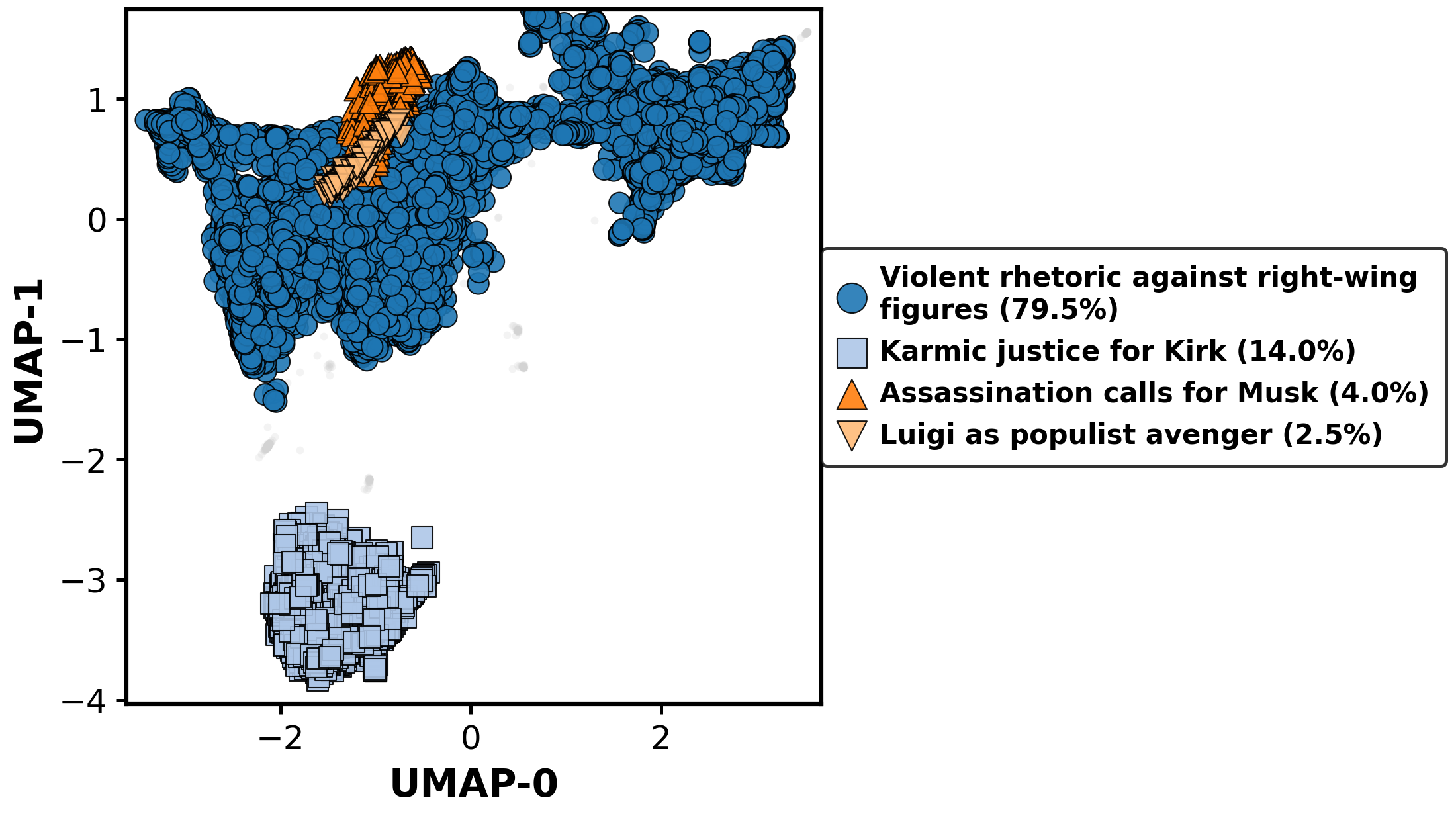}
    \caption{!takedown}
    \label{fig:takedown}
  \end{subfigure}
  \hspace{0.02\textwidth}
  \begin{subfigure}[t]{0.32\textwidth}
    \centering
    \includegraphics[width=\linewidth, height=3.5cm]{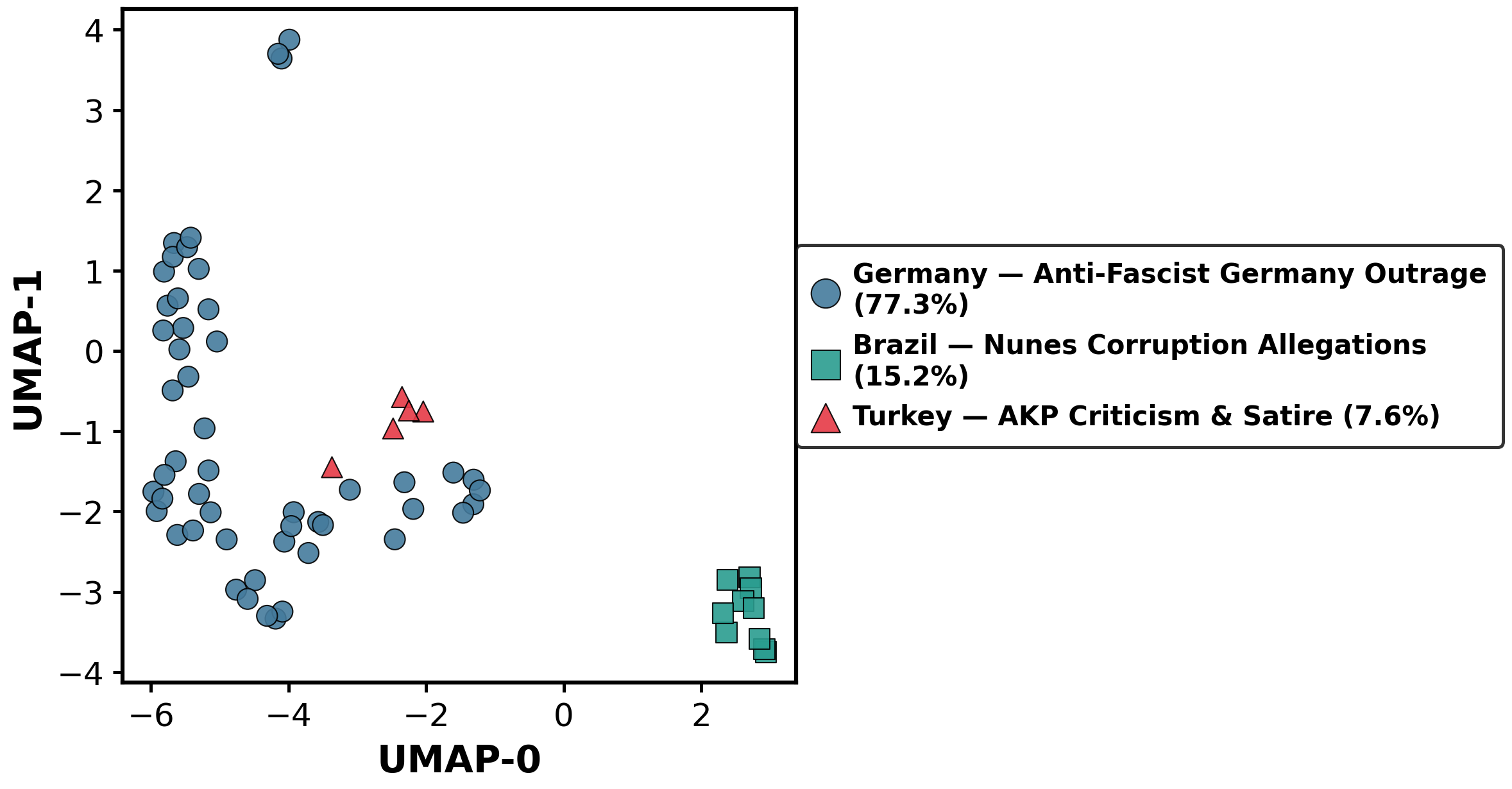}
    \caption{!hide (country-specific)}
    \label{fig:hidecountryclusters}
  \end{subfigure}
\end{center}
  \vspace{-2mm}
  \caption{Visualizations of clusters for different labels (first two components of UMAP)}
  \label{fig:six_umap_plots}
  \vspace{-3mm}
\end{figure*}

The goal of this analysis is to obtain robust clusters for each moderation label as mentioned in Section~\ref{Sec: Transparent}.

\noindent
\subsection{UMAP dimensions and HDBSCAN clustering:}  We performed hyperparameter tuning for HDBSCAN clustering to ensure coherent and meaningful groupings of posts. High-dimensional multimodal embeddings \new{(2048-d) were obtained using \texttt{Qwen3-VL-Embedding-2B}~\cite{li2026qwen3}}. 
\neww{This model allows encoding text, multiple images, and video into a single joint embedding vector. The embedding model follows the \texttt{Qwen3-VL} chat template, in which the multimodal instance- text, image(s), and video in any combination is passed as a single user turn, and the final hidden state at the sequence's last token (via last-token pooling) is taken as the fixed-length (2048-d) representation of the entire post.}
These embeddings were then projected into a low-dimensional space using UMAP. We explored a range of $n\_neighbors$ and $n\_components$, using Trustworthiness score to assess the preservation of local structure. 
Based on these evaluations, $n\_components = 5$ was selected as a reasonable trade-off for dimensionality reduction, and this setting was used for subsequent HDBSCAN clustering.
Since hyperparameter optimisation requires running UMAP and HDBSCAN
across dozens to hundreds of configurations, processing the full corpus
at each trial would be both computationally prohibitive and memory
intensive.
\new{Therefore, when a label corpus exceeded 300,000 posts, a uniform random subsample
of that size was drawn.}

Clusters were obtained from the UMAP embeddings using
HDBSCAN~\cite{campello2013density}, which additionally produces a soft
membership probability for each post reflecting its confidence of
belonging to its assigned cluster.
\new{Rather than exhaustive grid search, combinations of UMAP
\texttt{n\_neighbors}, HDBSCAN \texttt{min\_cluster\_size}, and
\texttt{min\_samples} were evaluated via Bayesian optimisation using
Optuna with a Tree-structured Parzen Estimator
(TPE) sampler~\cite{10.1145/3292500.3330701}.
Search ranges were adapted to corpus size.} Each configuration was assessed using the Density-Based Clustering
Validation (DBCV) score~\cite{doi:10.1137/1.9781611973440.96}, which measures
intra-cluster density relative to inter-cluster separation.
The configuration with the highest DBCV score was selected for final
clustering.
Table~\ref{tab:hyperparam} reports the selected hyperparameters, and DBCV scores for all labels.

While quantitative metrics such as DBCV provide an intrinsic measure of
cluster coherence, we also obtained qualitative descriptions and names for each cluster using the LLM-based pipeline described
in the following section.

\vspace{2mm}
\noindent\new{\subsection{Interpreting clusters}
To characterise the thematic content of each cluster, we employ a two-stage automated pipeline: we first draw a sample from each cluster, then pass the sampled
posts to a vision-language model to generate a concise thematic
description. Later, we also validate our pipeline with human evaluation on a subset of clusters.}

\new{\textbf{Sampling for LLM summarisation: } Since clusters vary substantially in size --- from hundreds to tens of thousands of posts --- passing all posts to the model is neither feasible \cite{ghali2025beyondwordsneedagenticgenerative} nor necessary: large clusters exceed any model's context window, and the semantic redundancy inherent in dense clusters means a statistically sufficient sample captures the same thematic signal at a fraction of the inference cost. For each cluster, a statistically justified sample was drawn using the Cochran~\cite{cochran1977sampling} finite-population-corrected formula:}

\new{\begin{equation}
    n = \frac{n_0}{1 + \frac{n_0 - 1}{N}}, \qquad
    n_0 = \frac{z^2\, p\,(1-p)}{e^2}
    \label{eq:cochran}
\end{equation}}

\noindent \new{where $n$ is the sample size, $z=1.96$ is the Z-score corresponding to a
95\% confidence level, $p=0.5$ is the estimated proportion of the
population (set to 0.5 for maximum sample size), and $e=0.05$ is the margin
of error.
Posts were sampled without replacement, weighted by their HDBSCAN soft membership probability, ensuring that sampled posts span the full thematic range of each cluster rather than concentrating on its dense core.}

\vspace{2mm}
\noindent\new{\textbf{LLM Summarisation.}
Sampled posts were passed to \texttt{Qwen3-VL-32B-Instruct}\footnote{\url{https://huggingface.co/Qwen/Qwen3-VL-32B-Instruct}} using a
map-reduce summarisation strategy, which addresses the context length limit of a single-pass LLM summarisation. Posts were batched into chunks of 32, each summarised independently into
2--3 sentences (map), and the chunk-level summaries were merged into a single
1--2 sentence cluster description in a second pass (reduce).
Model inputs included post text and any attached media (images/video), enabling the model to reason over
visual content alongside text. All inference used greedy decoding (\texttt{do\_sample=False}) for
deterministic, reproducible outputs, with \texttt{max\_new\_tokens=256}
for both the map and reduce stages.
The prompts used at each stage are shown in
Tables~\ref{tab:prompt_chunk} and~\ref{tab:prompt_reduce}.}

\vspace{2mm}
\noindent\new{\textbf{Naming Clusters}
Cluster names were generated by presenting all cluster descriptions for
a given label to the model in a single prompt, instructing it to assign
short (2--5 word) names that distinguish each cluster from its siblings.
This joint generation prevents label collapse --- the tendency for
independently named clusters to converge on near-identical generic
phrases --- and produces a taxonomy that is both specific and internally
contrastive. Naming inference also used greedy decoding (\texttt{do\_sample=False})
with \texttt{max\_new\_tokens=512}. The prompt used for contrastive naming is shown in
Table~\ref{tab:prompt_naming}.
LLM-generated descriptions indicate that clusters generally capture
dominant content themes per label.
Some overlap between clusters exists, but the overall cluster
assignments are coherent and meaningful, allowing us to characterise
labeled content at a fine-grained level.
We provide cluster names and their sizes in Figures~\ref{fig:three_umap_plots} and~\ref{fig:six_umap_plots}.}

\vspace{2mm}
\noindent\new{\textbf{Human Validation of Cluster Names.}
To assess the quality of the LLM-generated cluster names, we conducted a human validation study using a custom annotation interface built with Streamlit. For each cluster, annotators were presented with the cluster name, its LLM-generated summary, and up to 10 representative posts, including any attached images and video, sampled using scheme as the LLM summarisation stage. To protect annotators from harmful content, images and video in sensitive label categories (e.g.\ \texttt{graphic-media}, \texttt{self-harm}) were avoided.}

\new{Annotators rated each cluster name on a five-point Likert scale reflecting the degree to which the name accurately and specifically describes the sampled posts: (5)~\textit{very appropriate} --- the name precisely captures the cluster's content; (4)~\textit{appropriate} --- mostly accurate with minor gaps; (3)~\textit{neutral} --- partially accurate; (2)~\textit{inappropriate} --- significant misalignment; (1)~\textit{very inappropriate} --- the name is misleading or irrelevant. For ratings of 3 or below, annotators were additionally asked to identify the primary issue (too broad, too narrow, misleading, unclear, or other) and optionally suggest an alternative name. Inter-annotator agreement was computed using Krippendorff's $\alpha$~\cite{Krippendorff2011ComputingKA} over the ordinal Likert ratings.}

\noindent\new{\textbf{Human Validation Results.}
Three annotators independently rated 31 clusters, yielding 93 ratings in total. Cluster names received a mean appropriateness rating of $\mu = 4.58$ (SD $= 0.63$, median $= 5$) out of 5. In total, 92.5\% of individual ratings were at the \textit{appropriate} or \textit{very appropriate} level ($\geq 4$), and 90.3\% of clusters had a mean rating $\geq 4$. No cluster received a rating of \textit{inappropriate} or below ($\leq 2$), and the remaining 7.5\% of ratings were \textit{neutral} (3), indicating minor misalignment in a small number of cases. Inter-annotator agreement was assessed using Krippendorff's $\alpha$ with an ordinal metric, yielding $\alpha = 0.686$, which falls within the \textit{substantial} agreement range~\cite{landis1977measurement}. Of all pairwise rating comparisons across annotators, 95.7\% differed by at most one point on the five-point scale, and 63.4\% were in exact agreement. The four pairs differing by two points all involved a rating of 3 versus 5, representing cases where one annotator found the name neutral while another found it very appropriate. No pair differed by more than two points. Taken together, these results indicate that the automated pipeline: HDBSCAN clustering followed by LLM-based summarisation and naming, produces cluster names that are largely meaningful and faithful to the underlying post content as perceived by human reviewers.}

\section{Action Labeled Contents and Spam}\label{Sec: appendix actions}

As discussed in Section~\ref{sec: BMS}, in addition to category labels, BMS also issues action labels such as \lhide{}, \lwarn{} and \ltakedown{}, which enforce moderation decisions for all users and override any individual default settings. We also analyze these labeled posts to understand the content and behaviors that trigger such actions.

\vspace{0.5 mm}
\noindent
\textbf{Takedown: }
Posts with the label \ltakedown{} are removed from Bluesky as they violate the platform's terms of service. 
Since there is no precise, content-based definition for this label, we analyze the labeled posts to characterize the types of content and behaviors that are taken down by the platform. 
\new{ The clusters (Figure~\ref{fig:takedown}) reveal content centered on extreme political hostility and violent celebration. Most of these posts focus on violent rhetoric and explicit calls for harm against right-wing figures and public personalities (clusters: Violent rhetoric against right-wing figures). Another cluster captures the widespread online glorification of the shooting of Charlie Kirk, marked by \textit{schadenfreude} and framing his death as karmic justice (cluster: Karmic justice for Kirk). One set of posts centers on \textit{Luigi Mangione} as a symbolic populist avenger, reflecting a broader culture of rage against the elites (cluster: Luigi as populist avenger). Together, these clusters reveal that content taken down on Bluesky is predominantly characterized by celebrations of real-world violence and explicit calls for harm against named individuals.}

\vspace{0.5 mm}
\noindent
\textbf{Hide: } 
When the \lhide{} label is applied to a post, it is hidden from user feeds. 
The clusters for the label \lhide{} capture content containing threats, harassment, or promotion of self-harm (\Cref{fig:hidei} in Appendix). 
Some clusters focus on violent threats against public figures, particularly Donald Trump and Elon Musk \new{(cluster: Trump-Musk assassination calls), as well as conspiracy narratives surrounding alleged Iranian assassination plots against Trump (cluster: Iranian assassination conspiracy).} 
\new{Another cluster captures a coordinated doxxing campaign targeting young tech professionals involved with DOGE.}

\vspace{0.5 mm}
\noindent
\textbf{Warn: } 
The \lwarn{} label is applied on a post to issue a warning on the content before it is shown. 
\new{The clusters (in Figure~\ref{fig:warnh}) reveal content centered on extreme political hostility and doxxing. One cluster (Violent anti-Trump/Musk rhetoric) captures violent anti-Trump and anti-Musk sentiment framing both as authoritarian threats, while other clusters contain posts about doxxing activity targeting DOGE employees through the circulation of leaked email lists.}

\vspace{0.5 mm}
\noindent
\textbf{Country-Specific Moderation Service (Hide): }These clusters (Figure~\ref{fig:hidecountryclusters} in the appendix) reveal the contents flagged by country-specific labelers. 
\newline
\textbf{Germany: } The labeled post clusters are associated with German political discourse and involve political criticism of Friedrich Merz, Sahra Wagenknecht, AfD, and Hubert Aiwanger by labeling them with vulgarities and accusing them of enabling far-right rhetoric.
\newline
\textbf{Turkey: } These \lhide{} posts focus on criticism and satire of the AKP government and its perceived authoritarian control over state institutions, including commentary on the cancellation of opposition leader İmamoğlu's diploma.
\newline
\textbf{Brazil: } The \lhide{} posts are highly specific to the São Paulo municipal elections, focusing on allegations connecting the chief of staff of Mayor Ricardo Nunes and the Primeiro Comando da Capital (PCC), a major organized crime syndicate.

These observations lead to a number of interesting implications for building effective content moderation systems.

\vspace{0.5 mm}
\noindent
\textbf{Spam: } \new{The \lspam{} label identifies unwanted content and behavior on Bluesky, ranging from bot-like activity and solicitation to coordinated inauthentic posting. However, most post records for this label were not retrieved and are effectively lost, significantly limiting the depth of analysis. The few available clusters reveal predominantly political and activist content, including urgent humanitarian appeals around ongoing geopolitical conflicts (cluster: Global crisis activism), organized resistance and protest against authoritarian regimes (cluster: Defiant protest against authoritarianism), and a blend of humanitarian advocacy with personal expression (cluster: Humanitarian appeals with personal joy).}

\end{document}